\documentclass[12pt,a4paper]{article}
\pdfoutput=1
\usepackage[numbers,sort&compress]{natbib}
\usepackage{epsfig, axodraw}
\usepackage{graphicx}  
\usepackage{amsmath}  
\usepackage{graphics}
\usepackage{afterpage}
\usepackage{array, amssymb}
\usepackage[flushleft]{threeparttable}
\usepackage{subcaption}
\usepackage{cancel}
\usepackage[mathlines]{lineno}
\usepackage{lettrine}
\usepackage[dvipsnames]{xcolor}
\usepackage{dcolumn, bm, epic, eepic, float} 
\usepackage{amsfonts}
\usepackage{bbm}
\usepackage{booktabs}
\usepackage{autobreak}
\usepackage{multirow}
\usepackage{longtable}
\newcommand{\nc}{\newcommand}
\nc{\beq}{\begin{equation}}  \nc{\eeq}{\end{equation}}
\nc{\bea}{\begin{eqnarray}}  \nc{\eea}{\end{eqnarray}}
\nc{\baa}{\begin{array}}     \nc{\eaa}{\end{array}}
\def\theequation{\arabic{section}.\arabic{equation}}

\input Zallman.fd

\usepackage{jheppub} 
\title{\boldmath Maximising CP Violation in Naturally\\[2mm] Aligned Two-Higgs Doublet Models}

\author[a]{Neda Darvishi,}
\author[b]{Apostolos Pilaftsis,} 
\author[c,d,e,f,g]{Jiang-Hao Yu}

\affiliation[a]{Department of Physics, Royal Holloway, University of London,\\ Egham, Surrey, TW20 0EX, United Kingdom}
\affiliation[b]{Department of Physics and Astronomy, University of Manchester,\\ Manchester, M13 9PL, United Kingdom}
\affiliation[c]{CAS Key Laboratory of Theoretical Physics, Institute of Theoretical Physics,\\ Chinese Academy of Sciences, Beijing 100190, China}
\affiliation[d]{School of Physical Sciences, University of Chinese Academy of Sciences,\\ Beijing 100049, P.R. China}
\affiliation[e]{Center for High Energy Physics, Peking University, Beijing 100871, China}
\affiliation[f]{School of Fundamental Physics and Mathematical Sciences, Hangzhou Institute \\ for Advanced Study, UCAS, Hangzhou 310024, China}
\affiliation[g]{International Centre for Theoretical Physics Asia-Pacific, Beijing/Hangzhou, China}

\emailAdd{neda.darvishi@rhul.ac.uk}
\emailAdd{apostolos.pilaftsis@manchester.ac.uk}
\emailAdd{jhyu@itp.ac.cn}

\bigskip
\abstract{
The Two-Higgs Doublet Model (2HDM) is a well-motivated theoretical framework that provides additional sources of CP Violation (CPV) beyond the Standard Model (SM). After studying the vacuum topology of a general (convex) 2HDM potential, we unambiguously identify three origins of CPV: (i)~Spontaneous CPV (SCPV), where the vacuum manifold has at least two degenerate CPV minima disconnected by domain walls, (ii)~Explicit CPV (ECPV) with one single CPV ground state, and (iii)~Mixed Spontaneous and Explicit CPV (MCPV), where the theory possesses more than one {\em non}-degenerate CPV local minimum. Most importantly, we define a novel complex parameter
$r_{\rm CP}$ whose norm and phase control the three different realisations of CPV, at least at the tree level. In all these scenarios, only two CPV phases can be made independent, as any third CPV parameter will always be constrained via the CP-odd tadpole condition. Since ECPV vanishes in 2HDMs where SM Higgs alignment is achieved naturally through accidental continuous symmetries, we analyse the possibility of maximising CPV through soft and explicit breaking of these symmetries. 
We~derive upper limits on key CPV parameters that quantify the degree of SM misalignment from constraints due to the non-observation of an electron Electric Dipole Moment (EDM). 
Finally, we delineate the CP-violating parameter space of the so-constrained naturally aligned 2HDMs that can further be probed at the CERN Large Hadron Collider (LHC). }

\begin{document}
\maketitle
\flushbottom


\setcounter{equation}{0}
\section{Introduction}

\lettrine{V}{iolation of CP in the Standard Model} (SM) arises predominantly from the Yukawa sector of the theory~\cite{Kobayashi:1973fv}, provided the anomalous generation of a CP-odd phase induced by instantons~\cite{tHooft:1976rip,PhysRevD.14.517} is ignored. In the SM scalar sector, CP Violation (CPV) is very much suppressed. In fact, it only enters the effective action via the Higgs--$Z$-boson mixing\- at three loops~\cite{Pilaftsis:1998pe}, e.g.~through the dimension-6 effective-field-theory operator
\begin{equation}
   \label{eq:SMCP}
O_{\rm CP}\, =\, \frac{\delta_{HZ}}{\Lambda^2} \big(\Phi^\dagger iD_\mu \Phi\big)\, \partial^\mu \big(\Phi^\dagger \Phi\big)\ +\ {\rm H.c.}\;, 
\end{equation}
where $\delta_{HZ}$ is a dimensionless real Wilson coefficient, $\Phi$ is the SM Higgs doublet, and $D_\mu$ is the gauge-covariant derivative acting on $\Phi$. As a consequence of the
suppressed CPV, SM fails to explain the Baryon Asymmetry of the Universe (BAU) \cite{AndreiDSakharov1991}. Likewise, the restricted SM field content proves inadequate to account for other cosmological observations like the origin of the Dark Matter (DM)~\cite{Zwicky:1933gu,Barbieri:2010mn,QUEST-DMC:2023nug}.

One of the simplest and rather promising extensions of the SM that enables one to address the above two cosmological problems is the so-called Two-Higgs-Doublet Model~(2HDM)
whose Higgs potential is described by two SU(2)$_L$ scalar iso-doublets, $\Phi_1$ and $\Phi_2$. The 2HDM 
can provide extra sources of CPV~\cite{Lee:1973iz,Branco:1980sz,Branco:1985aq,Weinberg:1990me} which can facilitate electro\-weak baryogenesis~\cite{Cohen:1993nk}, and possibly accommodate extra stable scalar states that play the role of the DM ~\cite{Barbieri:2010mn}. Nevertheless, given the great phenomenological success of the SM in describing fundamental interactions of elementary particles at colliders, the coupling strengths of the Higgs boson in the 2HDM, primarily to the electroweak~(EW) $W^\pm$ and $Z$ gauge bosons, must be very close to those predicted by the SM~\cite{ATLAS:2016neq,Palmer:2021gmo,ATLAS:2021upq}. 
This means that in any theory of New Physics, the candidate particle for the 125~GeV scalar resonance observed in 2012 at the CERN Large Hadron Collider (LHC) should have properties that are well aligned with those of the ordinary SM scalar boson $H_{\rm SM}$ in all its interactions with the SM fermions and the EW gauge bosons. Such SM alignments of interactions can be achieved either through fine-tuning~\cite{Chankowski:2000an,Gunion:2002zf,Ginzburg:2004vp,Carena:2013ooa}, or naturally through the imposition of symmetries ~\cite{Pilaftsis:2016erj,BhupalDev:2014bir,Darvishi:2019dbh}. Most notably, the $\mathbb{Z}_2$ and U(1) symmetries become instrumental in the Yukawa sector of the 2HDM~\cite{Glashow:1976nt,Peccei:1977ur}, in order to eliminate dangerous\- flavour-changing-neutral currents (FCNCs) of the SM-like Higgs boson to quarks at tree level.

Besides Natural Flavour Conservation (NFC) in the Yukawa sector, Natural SM-Higgs Alignment (NHAL) in the gauge sector can also be achieved through the imposition of ${\rm SU(2)}_L$-preserving continuous symmetries~\cite{Pilaftsis:2016erj,BhupalDev:2014bir,Darvishi:2020teg,Darvishi:2019ltl,Darvishi:2021txa,Pilaftsis:2022euw}, without the need to decouple the new heavy states~\cite{Chankowski:2000an,Gunion:2002zf,Ginzburg:2004vp,Carena:2013ooa}. This symmetry-enforced NHAL turns out to be independent of the ratio $\tan\beta = v_2/v_1$ of the $\Phi_{1,2}$ vacuum expectation values (VEVs), or the form and size of the bilinear mass terms added to the 2HDM potential. The only pheno\-menological constraint is the existence of a CP-even scalar state with mass equal to about~125 GeV which should be associated with the spontaneous symmetry breaking of the theory in the so-called Higgs basis~\cite{GEORGI197995,PhysRevD.19.945,Lavoura:1994fv,Botella:1994cs}. In~\cite{BhupalDev:2014bir}, three distinct continuous symmetry groups were found that enforce NHAL in the 2HDM at tree level: (i) the symplectic group ${\rm Sp}(4)$ acting on the 4D field space $\big(\Phi_1\,, \Phi_2\,, \widetilde{\Phi}_1\,, \widetilde{\Phi}_2\big)$, with~${\widetilde{\Phi}_{1\,(2)} = i\sigma_2 \widetilde{\Phi}^*_{1\,(2)}}$, and the two Higgs-Family~(HF) symmetries, (ii)~${\rm SU}(2)_{\rm HF}$ and (iii)~CP$\times {\rm SO}(2)_{\rm HF}$, operating on the 2D field-space~$\big(\Phi_1\,, \Phi_2\big)$. 

In the literature~\cite{Lee:1973iz,Branco:1980sz,Branco:1985aq,Weinberg:1990me}, CPV is broadly recognized to originate from two generic mechanisms: (a) CP is broken at the Lagrangian level, as exemplified by the SM with complex Yukawa couplings, and (b) CP is conserved at the Lagrangian level, but violated by the vacuum. In a scenario where CPV arises solely from the vacuum, it is important to note that this does not involve one single ground state, but at least two degenerate CP-violating vacua that are separated by domain walls~\cite{Battye:2011jj, Battye:2020sxy}. In particular, one cannot go from the one degenerate vacuum to the other one through SM gauge transformations, because the two CPV vacua live in disconnected regions of the vacuum manifold. Now, by turning on small explicit CP-odd phases, one can lift this degeneracy, but one still has two local vacua or minima that break CP. This last situation introduces a new and physically distinct scenario, 
termed here a Mixed CPV scenario, in which tunnelling from one vacuum to the other one can in principle take place through a first-order EW phase transition.
If the CP phases become large enough, then one gets a global CP-violating vacuum (up to overall SM gauge rotations), under the assumption that the tree-level 2HDM potential is convex. Consequently, one important finding of the present study is that there exists a key complex parameter $r_{\rm CP}$ of the general 2HDM potential, as defined in~\eqref{eq:rCP},  whose absolute value $|r_{\rm CP}|$ and its phase $\phi_{\rm CP}$ enables one to unambiguously characterise the three possible realisations of CPV: (i)~Spontaneous CP Violation (SCPV),  (ii)~Explicit CP Violation (ECPV) and (iii)~Mixed Spontaneous and Explicit CP Violation (MCPV).

As opposed to an aligned scenario that heavily relies on fine-tuning in the Higgs basis, NHAL symmetries\- give rise to 2HDM potentials whose quartic couplings are all real~\cite{BhupalDev:2014bir,Darvishi:2019ltl,Darvishi:2020teg}. As a consequence of the latter, one is confronted with the undesirable feature that all explicit sources of CPV vanish~\cite{Grzadkowski:2014ada}. Nevertheless, there is one additional option that we wish to study in detail in this work. This is the possibility of SCPV for real Higgs potentials. Such a realisation can only occur within 2HDM potentials that realise the aforementioned third symmetry, CP$\times {\rm SO(2)}_{\rm HF}$ as well as two custodial symmetries $\text{CP}\times\text{O}(4)$ and $\text{O}(2)\times \text{O}(3)$, up to softly broken CP-even mass terms.

In this paper we will also analyze another alternative for generating sizeable CPV in the scalar sector, whilst being close to NHAL. Specifically, we introduce small departures from NHAL through explicit CP-violating quartic couplings or through a complex bilinear mass term, $m^2_{12}\Phi^\dagger_1\Phi_2$, in the 2HDM potential. As a result, the larger the deviation from NHAL is, the larger CPV gets realized. One of the central goals of the paper is therefore to maximise~CPV, while maintaining agreement with the current LHC data, along with the strict upper limits on the electron EDM. Consequently, all misalignment directions that we classify in this second option will break CPV explicitly as well. 

The rest of the paper is laid out as follows. In Section~\ref{sec:CPV2HDM}, we discuss the vacuum topo\-logy of a general CP-violating 2HDM potential.
In particular, after taking into account all options in which CP can be broken, we identify from the profile of the vacuum manifold three distinct scenarios: (i)~SCPV, (ii)~ECPV and (iii) MCPV. 
The analytic results of scalar masses and Higgs couplings that will be relevant to our phenomenological analysis are given in Appendix~\ref{app:2.2}. In Section~\ref{sec:CSNAL}, we briefly review the classification of 
2HDM scenarios that realise approximate NHAL, whilst a weak-basis independent condition for SM Higgs alignment is presented in~Appendix~\ref{app:HAL}. In Section~\ref{sec:SO2HDM}, we analyse SO(2)- and SO(4)-symmetric 2HDMs that depart from the NHAL symmetries, ${\rm CP}\times {\rm SO}(2)$ and ${\rm CP} \times  {\rm O}(4)$, by breaking CP explicitly. In addition, the LHC constraints on a few representative scenarios are considered, as well as the impact of the electron EDM limits on the maximally allowed size of CPV. Pertinent analytic results of the electron EDM at two loops are given in Appendix~\ref{app:eEDM}. Our conclusions are summarised in Section~\ref{sec:Concl}.

\setcounter{equation}{0}
\section{Vacuum Topology of the CP-violating 2HDM}\label{sec:CPV2HDM}

In this section, we will first review the basic structure of 
the general CP-violating 2HDM potential, as well as investigate certain aspects of its vacuum topology. We will then present the Higgs mass spectrum and all Higgs couplings relevant to our study. Our notation and conventions for the CP-violating 2HDM potential will follow to a good extent those introduced in~\cite{Pilaftsis:1999qt}. 

Our starting point is the scalar potential $\mathcal{V}$ of the 2HDM which may be expressed in terms of the two Higgs doublets $\Phi_1$ and $\Phi_2$ as follows:
\begin{align}
   \label{eq:Vpot}
\mathcal{V} &= -\frac12 \bigg[ m_{11}^2 |\Phi_1|^2 +m_{22}^2 |\Phi_2|^2 + \Big(m_{12}^2 \Phi_1^\dagger \Phi_2+ {\rm H.c.} \Big)\bigg] + { \lambda_1} |\Phi_1|^4 +{ \lambda_2} |\Phi_2|^4
	 \\
	 &+ \lambda_3 |\Phi_1|^2 |\Phi_2|^2+\lambda_4  |\Phi_1^\dagger \Phi_2|^2 + \bigg( \frac12 \lambda_5(\Phi_1^\dagger \Phi_2)^2+ \lambda_6 (\Phi_1^\dagger \Phi_2)|\Phi_1|^2+ \lambda_7 (\Phi_1^\dagger \Phi_2)|\Phi_2|^2+ {\rm H.c.} \bigg).
\nonumber
\end{align}
Note that the parameters, $m_{12}^2$ and $\lambda_{5,6,7}$, are complex, whereas the remaining parameters, $m^2_{11}$, $m^2_{22}$ and $\lambda_{1,2,3,4}$, are real. Our next step is to determine the ground state of the Higgs potential, and so study the topology of its vacuum manifold in its CP-odd phase or CP-odd scalar field direction. 

We start by considering the linear decompositions of the Higgs doublets,
\begin{equation}
  \label{Phi12}
\Phi_1\ =\ \left( \begin{array}{c}
\phi^+_1 \\ \frac{1}{\sqrt{2}}\, ( v_1\, +\, \phi_1\, +\, ia_1)
\end{array} \right)\, ,\qquad
\Phi_2\ =\ e^{i\xi}\, \left( \begin{array}{c}
\phi^+_2 \\  \frac{1}{\sqrt{2}}\, ( v_2 \, +\, \phi_2\, +\, ia_2 )
 \end{array} \right)\, ,
\end{equation}
where $v_1$ and $v_2$ are the moduli of the vacuum expectation values
(VEVs) of the Higgs doublets and $\xi$ is their relative phase.  
 Here, we have adopted a weak basis in which VEVs $v_1,v_2$ and the quantum fluctuations $\phi_1,\phi_2$
hold the same phase. 
The minimization conditions resulting from the potential give rise to the following relations~\cite{Pilaftsis:1999qt}:
\begin{eqnarray}
T_{\phi_1} &\equiv & \Big<\frac{\partial \mathcal{V}}{\partial
  \phi_1}\Big>\ =\ 
\frac12 v\, c_\beta\, \Big[ -  m_{11}^2\ -\  {\rm Re} (m^2_{12}e^{i\xi})\,
  \tan\beta\, +\,\Big(\, 2 \lambda_1 c^2_\beta\, +\,
 \, (\lambda_3 + \lambda_4) s^2_\beta\, \nonumber\\
&&+\, {\rm Re}(\lambda_5 e^{2i\xi})s^2_\beta\, +\, 
3\, {\rm Re}(\lambda_6 e^{i\xi})s_\beta c_\beta\, +\,
 \, {\rm Re}(\lambda_7 e^{i\xi})s^2_\beta\tan\beta\, \Big) v^2\,
  \Big]\, ,  \label{Tphi1} \\
T_{\phi_2} &\equiv & \Big<\frac{\partial \mathcal{V}}{\partial
  \phi_2}\Big>\ =\ 
\frac12 v\, s_\beta\, \Big[\, -   m_{22}^2\ -\   {\rm Re} (m^2_{12}e^{i\xi})\,
  \cot\beta\, +\,\Big(\, 2 \lambda_2 s^2_\beta\, +\,
  \, (\lambda_3 + \lambda_4) c^2_\beta\, \nonumber\\
&&+\, {\rm Re}(\lambda_5 e^{2i\xi})c^2_\beta\, +\, 
 \, {\rm Re}(\lambda_6 e^{i\xi})c^2_\beta\cot\beta\, +\,
3\, {\rm Re}(\lambda_7 e^{i\xi})s_\beta c_\beta\, \Big) v^2
  \Big]\, ,  \label{Tphi2}
\end{eqnarray}  
\begin{eqnarray}
T_{a_1} &\equiv & \Big<\frac{\partial \mathcal{V}}{\partial a_1}\Big>
\ =\ \frac12 v\, s_\beta\, \Big( - {\rm Im} (m^2_{12}e^{i\xi})\, +\, 
{\rm Im}(\lambda_5 e^{2i\xi})\, v^2 s_\beta c_\beta\, 
+\, {\rm Im}(\lambda_6 e^{i\xi}) v^2 c^2_\beta\, \nonumber\\
&&+\,  {\rm Im}(\lambda_7 e^{i\xi}) v^2 s^2_\beta\, \Big)~,  \label{Ta1} \\
T_{a_2} &\equiv & \Big<\frac{\partial \mathcal{V}}{\partial a_2}\Big>
\ =\ -\, \frac12 v\, c_\beta\, \Big( -  {\rm Im} (m^2_{12}e^{i\xi})\, +\, 
{\rm Im}(\lambda_5 e^{2i\xi})\, v^2 s_\beta c_\beta\, +\, 
  {\rm Im}(\lambda_6 e^{i\xi}) v^2 c^2_\beta\, \nonumber\\
&& +\,  {\rm Im}(\lambda_7 e^{i\xi}) v^2 s^2_\beta\, \Big)~,   \label{Ta2}
\end{eqnarray}
where $s_x\equiv \sin x$,  $c_x\equiv  \cos x$, $\tan\beta =  v_2/v_1$
and $v  = \sqrt{v^2_1 + v^2_2}\,$ is the SM VEV. Moreover, the variations of the scalar potential $\mathcal{V}$ with respect to $\phi^+_{1,2}$ vanish identically in the absence of a non-zero charged vacuum according to the linear expansions~of $\Phi_{1,2}$ in~\eqref{Phi12}. As shown in~\cite{Branco:1980sz,Sher:1988mj}, a charged vacuum cannot co-exist with other neutral vacua at tree level. This~is what is tacitly assumed in this work, along with the bounded-from-below convexity conditions for a general 2HDM potential~\cite{Bahl:2022lio,Song:2023tdg}.

The neutral would-be Goldstone boson~$G$ associated with the longitudinal degree of the $Z$ boson and its orthogonal CP-odd quantum fluctuation~$a$ may be expressed in terms of the CP-odd scalars~$a_{1,2}$ as
\begin{equation} 
   \label{eq:rotGa}
 G\, =\, c_\beta a_1 + s_\beta a_2\;, \qquad a\, =\, -s_\beta a_1 + c_\beta a_2\;.
\end{equation}
By employing the above orthogonal transformation~\eqref{eq:rotGa} involving the mixing angle~$\beta$, we~find that the $G$-tadpole parameter $T_G$ vanishes identically, as it should be~\cite{Pilaftsis:1998pe}, whilst the tadpole parameter $T_a$ pertinent to the CP-odd field~$a$ becomes~\cite{Pilaftsis:1999qt}
\begin{eqnarray}  
  \label{eq:Ta}
T_a &\equiv& \Big<\frac{\partial {\mathcal{V}}}{\partial a}
\Big>\ =\ -\, \frac{1}{2}\,v\, \Big( -{\rm Im} (m^2_{12}e^{i\xi})\, +\,
{\rm Im}(\lambda_5 e^{2i\xi})\, v^2 s_\beta c_\beta\, +\, 
\frac{1}{2}\, {\rm Im}(\lambda_6 e^{i\xi})\, v^2 c^2_\beta\, \nonumber\\
&& +\, \frac{1}{2}\, {\rm Im}(\lambda_7 e^{i\xi})\, v^2 s^2_\beta \, \Big)~.
\end{eqnarray}
We note that in the CP-invariant limit of the complete theory, $T_a$ tends to zero and the phase $\xi$ takes on the CP-conserving values~0 or $\pi$, with $\xi \in (-\pi\,,\pi ]$. In a general 2HDM with Higgs-sector CPV, the phase $\xi$ already occurs in the Born approximation. But taking all scalar potential mass parameters and quartic couplings to be real at the tree level, a non-zero value of the phase $\xi$ could be generated radiatively via loop effects if CPV exists in other sectors of the theory~\cite{Pilaftsis:1998pe}. For instance, this possibility arises in the so-called Minimal Supersymmetric Standard Model (MSSM) with explicit CPV~\cite{Pilaftsis:1998dd,Pilaftsis:1999qt}. 

By analogy, one may determine the physical $H^\pm$ boson and the would-be Goldstone boson $G^\pm$ under the same orthogonal transformation:
\begin{equation}
 G^{\pm} = c_\beta \phi_1^{\pm} + s_\beta \phi_2^{\pm}\,, \qquad H^{\pm} = -s_\beta\phi_1^{\pm} + c_\beta \phi_2^{\pm}\;.
\end{equation}
The charged-Higgs-boson mass is then given as
\beq
M^2_{H^\pm} = \frac{1}{2 s_\beta c_\beta}\, 
\Big\{ {\rm Re}( m^2_{12} e^{i\xi}) - \Big[\,
 \Big(\lambda_4 +
{\rm Re}(\lambda_5 e^{2i\xi})\Big) s_\beta c_\beta +
 \, {\rm Re}(\lambda_6 e^{i\xi}) c^2_\beta +  \, {\rm Re}(\lambda_7 e^{i\xi}) s^2_\beta \Big]\,v^2\,
\Big\}\, .
 \label{MHp}
\eeq
We require that the above expression for $M^2_{H^\pm}/v^2$ be positive at a minimum of the 2HDM potential, as expected to be the case for a neutral vacuum.

Now, from the minimization condition~\eqref{eq:Ta} and after
introducing the two phases,
\begin{align}
\varphi_5 \equiv \text{arg}(\lambda_5) \,=\, \tan^{-1}\left(\frac{\mathrm{Im}(\lambda_5)}{\mathrm{Re}(\lambda_5)}\right),\quad 
\varphi_{12} \equiv \tan^{-1}\left(\frac{\text{Im}\,\big(m^2_{12}  -\, 
                 \lambda_6\,c^2_\beta v^2\, -\, \lambda_7\,s^2_\beta v^2\big)}{\text{Re}\,\big(m^2_{12}  -\, 
                 \lambda_6\,c^2_\beta v^2\, -\, \lambda_7\,s^2_\beta v^2\big)}\right),
\end{align}
the following general constraint on the CP phases of the 2HDM
potential can be deduced: 
\begin{align} 
&\sin(2\xi+\varphi_5)\,=\,\frac{\big{\vert} m_{12}^2 -\, 
                 \lambda_6\,c^2_\beta v^2\, -\, \lambda_7\,s^2_\beta v^2\big{\vert}}{
                 \big{\vert}\lambda_5\big{\vert} s_\beta c_\beta 
                 v^2 } \sin(\xi+\varphi_{12})\,. 
\end{align}
Equivalently, the above constraint can be re-expressed as
 \begin{align}
{ 2\, |r_{\rm CP}| \sin(\xi+\varphi_{12})\: -\: \sin(2\xi+\varphi_5) }\: 
=\: 0 \,,
\label{MaxCP}
\end{align}
with 
\begin{equation}
  \label{eq:rCP}
r_{\rm CP}\: \equiv\: \frac{m_{12}^2 -\,
                 \lambda_6\,c^2_\beta v^2\,  -\, \lambda_7\,s^2_\beta v^2}{\lambda_5s_{2\beta}\, v^2 }\ .  
\end{equation}
Observe that by writing $r_{\rm CP} = |r_{\rm CP}|\,e^{i\phi_{\rm CP}}$, 
we derive the key relation,
\begin{equation}
  \label{eq:phiCP}
\phi_{\rm CP}\: =\: \varphi_{12}\, -\,
\varphi_5\:,
\end{equation}
with $\phi_{\rm CP} \in (-\pi\,, \pi]$. As we 
will see below, the values of $|r_{\rm CP}|$ and its phase $\phi_{\rm CP}$ play an instrumental role, as they allow us to identify the
different realisations of CPV in the 2HDM potential.
 
To simplify matters, we consider the weak-basis choice:
\begin{equation}
   \label{eq:MNbasis}
\lambda_6\, =\, \lambda_7,\qquad {\rm Im}\lambda_5\, =\, 0\,,\qquad 
\lambda_5\, \ge\, 0,   
\end{equation}
from which one uniquely gets $\varphi_5 = 0$. This in turn implies that $\phi_{\rm CP} = \varphi_{12}$ and
\begin{equation}
  \label{eq:rCPabs}
|r_{\rm CP}|\, =\, \frac{\big|m_{12}^2\, -\,
                 \lambda_6\,v^2\big|}{\lambda_5 s_{2\beta}\,v^2 }\; .
\end{equation}
As shown in~~\cite{Maniatis:2011qu}, such a weak-basis choice in~\eqref{eq:MNbasis} is always possible by taking advantage of the freedom of an SU(2) reparameterisation of the two Higgs doublets, $\Phi_1$ and~$\Phi_2$, which modifies the 2HDM potential accordingly. In fact, this choice of weak basis uniquely fixes the scalar potential, by rendering
the existence of possible accidental symmetries
manifest~\cite{Pilaftsis:1999qt,Battye:2011jj}. Hence, in the same
weak basis, we may unambiguously identify the three origins of CP
violation in a general 2HDM:

\begin{itemize}

\item[{\bf (i)}] {\bf Spontaneous CP Violation (SCPV)}~\cite{Lee:1973iz,Branco:1980sz,Branco:1985aq}, where only $\xi$ is non-zero and all parameters in the 2HDM potential are real,
i.e.~${\rm Im}(\lambda_6) = {\rm Im}(\lambda_7)=0$ and
${\rm Im}(m^2_{12}) =0$, so $\phi_{\rm CP} = 0$ or~$\pi$. As~can be
deduced from the constraint in~\eqref{MaxCP}, one must have
$0<|r_{\rm CP}| < 1$.  In this case, the vacuum manifold
$\mathcal{M}$ of the 2HDM potential contains two disconnected
degenerate neutral vacua,
i.e.~$\mathcal{M}=\{(v_1,\,v_2 e^{i\xi}),\,(v_1,\,v_2
e^{-i\xi})\}$, which cannot be transformed into one another by SM gauge transformations and hence may lead to the formation of domain walls~\cite{Battye:2011jj,Chen:2020soj,Battye:2020sxy}.

\item[{\bf (ii)}] {\bf Explicit CP Violation (ECPV)}~\cite{Weinberg:1990me,Pilaftsis:1999qt}, where  i.e.~${\rm Im}(\lambda_6) = {\rm Im}(\lambda_7)\neq 0$ and ${\rm Im}(m^2_{12}) \neq 0$. As we will analyse below, in its purest version, this scenario occurs when $|r_{\rm CP}| > 1$. In this case, only one neutral global
minimum exists, provided the tree-level 2HDM potential is bounded from below. According to the CP-odd tadpole conditions~\eqref{eq:Ta} or~\eqref{MaxCP}, when 
\begin{align}
{\rm Im} (m^2_{12})={\rm Im}(\lambda_6) v^2,
\end{align}
this scenario may include possibilities for which the CP-odd phase $\xi$ takes the CP-conserving values: 0 or~$\pi$.

\item[{\bf (iii)}]{\bf  Mixed Spontaneous and Explicit CP Violation (MCPV)},
 with three
  non-zero physical CP phases: 
$ \xi \neq 0,\, {\rm Im}(\lambda_6) = {\rm Im} (\lambda_7) \neq 0$ and
${\rm Im} (m^2_{12}) \neq 0$. Most importantly, MCPV is characterised 
by two local CP-violating minima of the 2HDM potential $\mathcal{V}$,
such that one is deeper than the other at two different phases $\xi
=\xi_{1,2}$, i.e.~$\mathcal{V}(v_1,v_2 
e^{i\xi_1}) < \mathcal{V}(v_1,v_2 e^{i\xi_2})$. 

\end{itemize} 

\begin{figure}[t]
\centering
\includegraphics[width=0.485\textwidth]{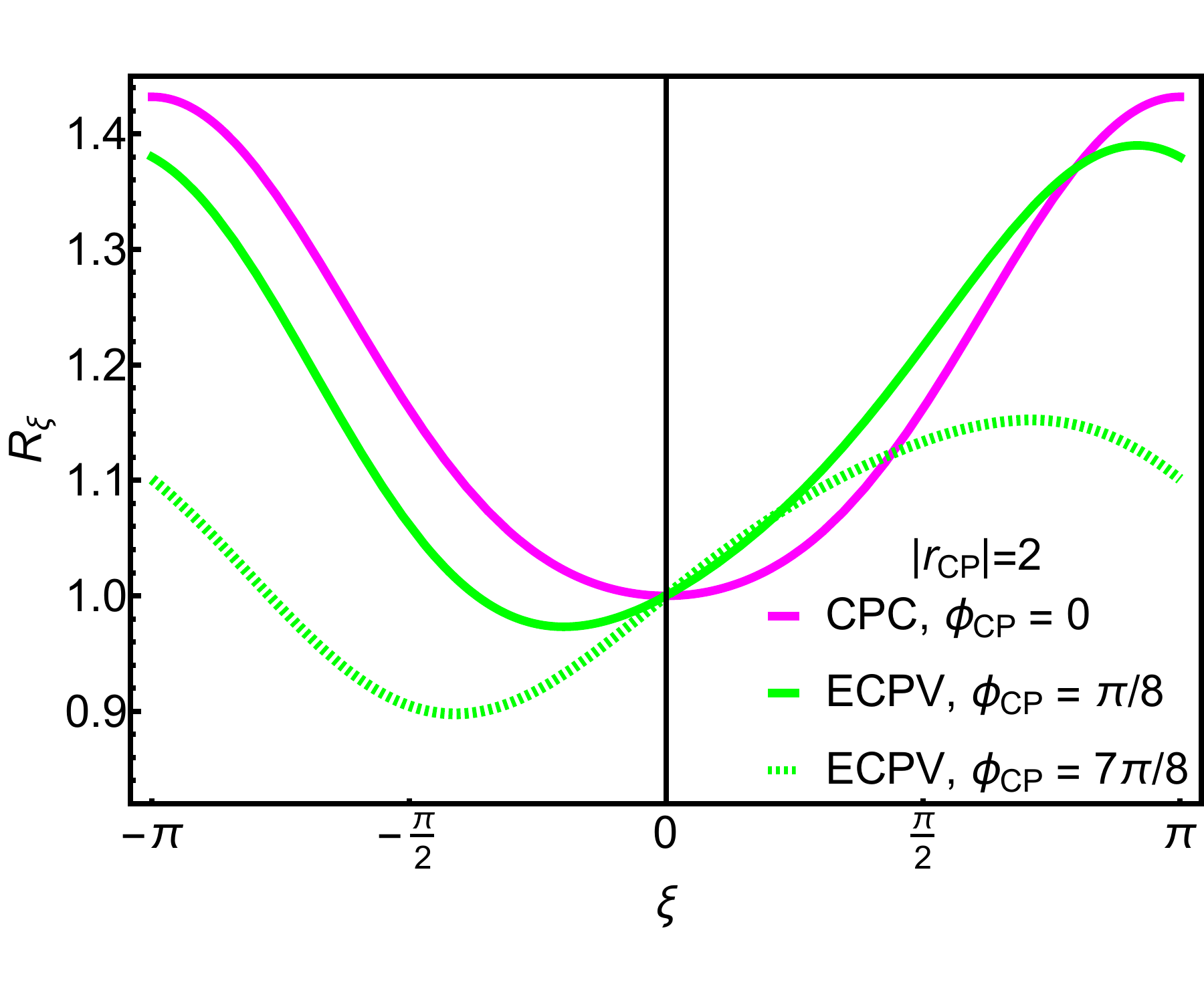}
\includegraphics[width=0.485\textwidth]{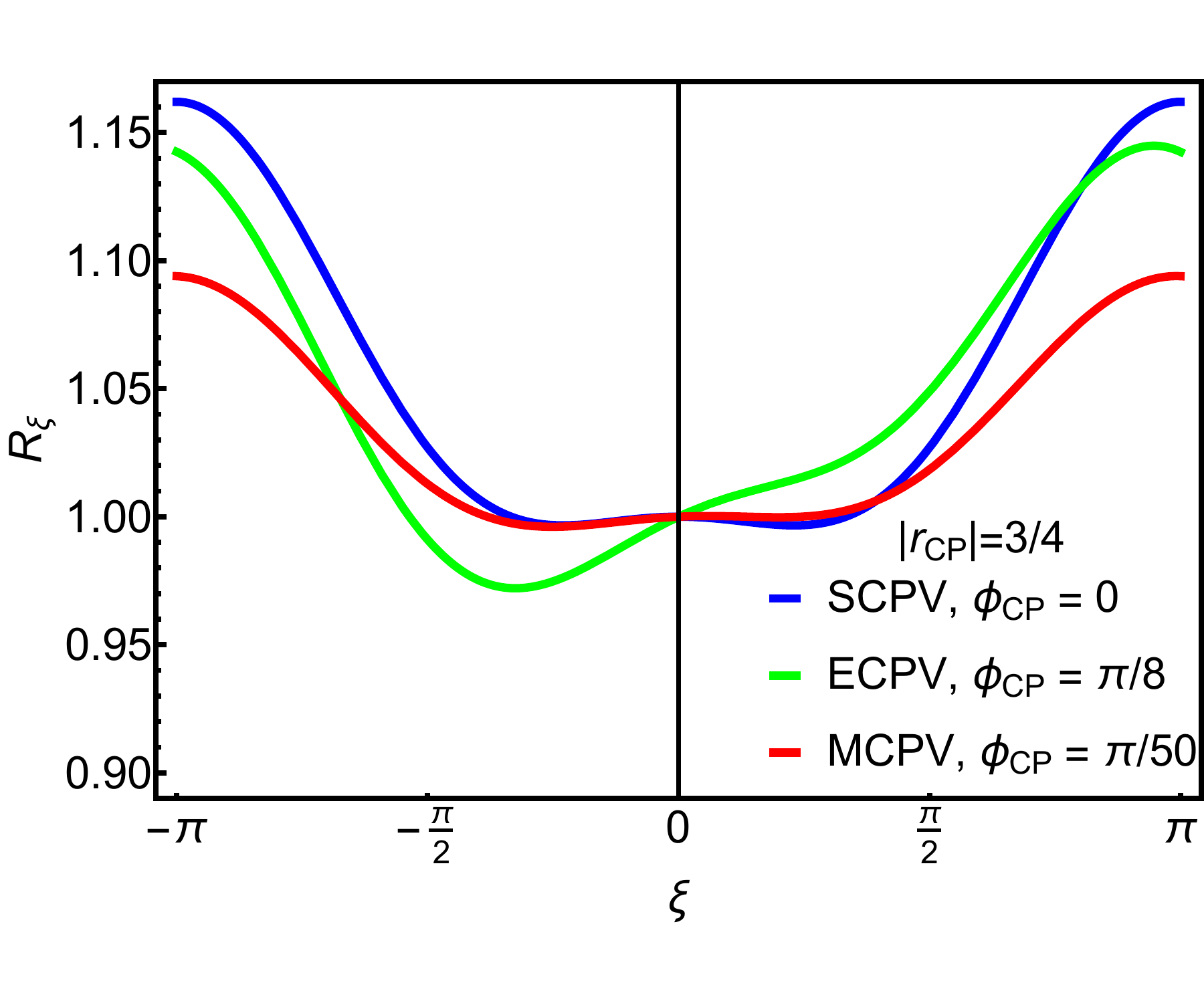}
\includegraphics[width=0.485\textwidth]{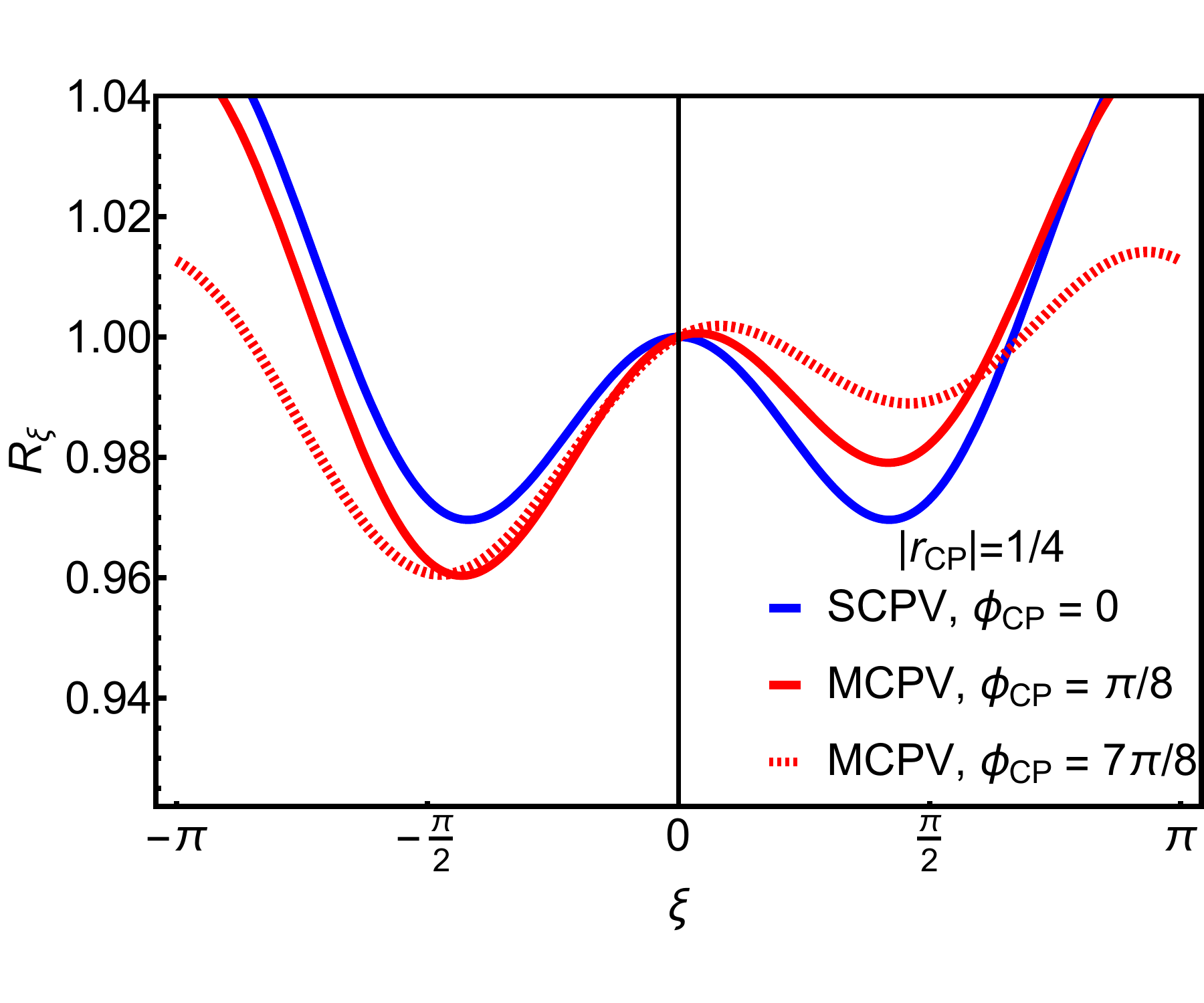}
\includegraphics[width=0.485\textwidth]{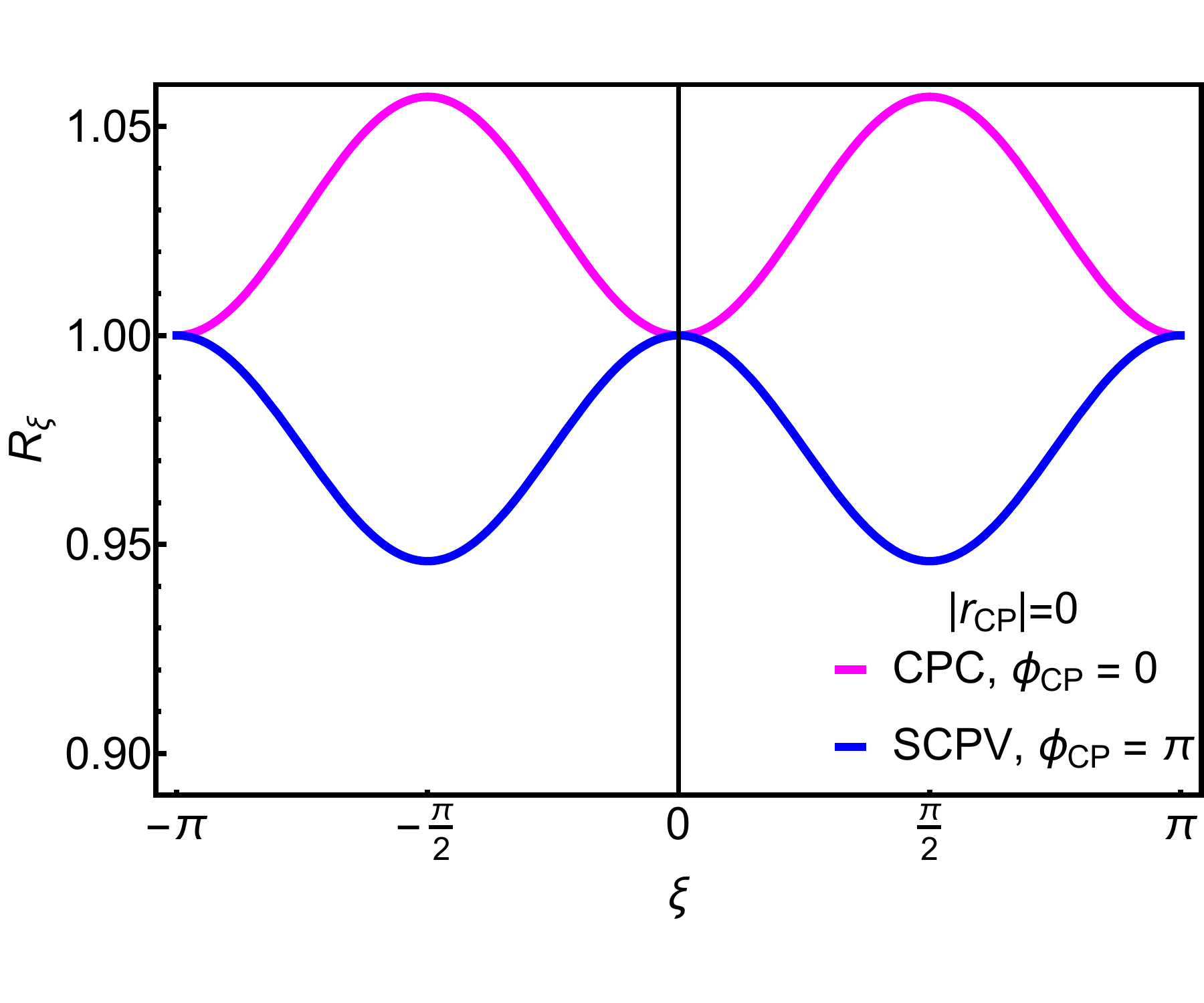}
\caption{\textit{The ratio $R_\xi\equiv \mathcal{V}(v_1,v_2 e^{i\xi})/|\mathcal{V}(v_1,v_2)|$ versus the phase $\xi$ is shown for selective values of $|r_{\rm CP}|$ and $\phi_{\rm CP} = \varphi_{12}$, and $\tan\beta = 2$. Shown are the different vacuum topologies of the 2HDM potential for CPC (magenta), ECPV (green), SCPV (blue) and MCPV (red).}}
\label{fig:Rxi}
\end{figure}

In order to illustrate the above three realisations of CPV that can take place in the general 2HDM potential of~\eqref{eq:Vpot}, we define the following ratio of scalar potentials:
\begin{align}
   \label{eq:Rxi}
R_\xi\: \equiv\: \frac{\mathcal{V}_\xi}{ |\mathcal{V}_{\xi=0}| }\ =\ \text{sgn}\,D\: +\: \frac{N}{ |D|}\; ,
\end{align}
where $\text{sgn}\,x$ is the standard sign function which is $+1$ for positive arguments of $x$ and $-1$ for negative values of its argument. In writing the RHS of \eqref{eq:Rxi}, we have re-expressed the scalar potential as: $\mathcal{V}_\xi \equiv \mathcal{V}(v_1,v_2\, e^{i\xi}) = \frac{v^2}{32} \big(D + N(\xi)\big)$, such that $\mathcal{V}_{\xi=0} = \frac{v^2}{32} D$
and~${N=N(\xi)}$ contains all non-trivial $\xi$-dependence, with $N(0) = 0$. The VEVs $v_{1,2}$ satisfy the CP-even tadpole conditions $T_{\phi_1} = T_{\phi_2} = T_a = 0$, but only for the global minimum of the scalar potential at $\xi =\xi_{\rm min}$. However, for the sake of illustration,  we eliminate the mass parameters $m^2_{11}$ and $m^2_{22}$ using~\eqref{Tphi1} and~\eqref{Tphi2}, respectively, in favour of the other parameters of the scalar potential, while keeping fixed the CP-odd phase $\xi$ to the reference value:~{$\xi=0$}. In~this~way, an adequate representation of the vacuum topology of the 2HDM potential in~\eqref{eq:Vpot} can be obtained along the $\xi$ direction. In detail, in the weak basis: ${\rm Im}\,\lambda_5 = 0$ and $\lambda_6=\lambda_7$, we have
\begin{align}
\label{eq:Vxi}
\mathcal{V}_\xi\ =&\ { v^2\over 32}\bigg\{v^2 \big( 6 \lambda_1 + \lambda_3 + \lambda_4\big) -
    4 \big(m_{11}^2 + m_{22}^2\big)- 4 \big( m_{11}^2 - m_{22}^2\big) c_{2\beta} 
     \nonumber   \\&\ +  
    v^2 \big(2 \lambda_1 - \lambda_3 - \lambda_4\big) c_{4\beta}
-2 s_{2\beta} \Big[ 4 \Big({\rm Re}(m_{12}^2) - {\rm Re}(\lambda_6) v^2\Big) \cos\xi 
 \nonumber   \\&\ - \lambda_5 s_{2\beta} v^2 \cos 2 \xi  - 4 \Big({\rm Im}(m_{12}^2) -{\rm Im}(\lambda_6) v^2 \Big) \sin\xi\Big]\bigg\}\;.
\end{align}
With the help of the analytic expression in~\eqref{eq:Vxi}, the quantities $N$ and $D$ that appear in~\eqref{eq:Rxi} are 
found to be
\begin{align}
  \label{eq:Nparam}
N\, =&\ 8 s_{2\beta} \Big[
\big(1-\cos\xi \big)\,\Big({\rm Re}(m_{12}^2)-{\rm Re}(\lambda_6)v^2\Big)\,+\, \sin \xi\,\Big( {\rm Im}(m_{12}^2)-{\rm Im}(\lambda_6)v^2 \Big)\nonumber\\
&\:-\: \sin^2\xi\, \lambda_5 s_\beta c_\beta v^2 \Big]\nonumber\\ =&\
4\,|\lambda_5| s_{2\beta}^2 v^2\,\Big[\, 2\, |r_{\rm CP}| \Big(\cos\varphi_{12} - \cos(\xi +\varphi_{12}) \Big)\: -\: \text{sgn}(\lambda_5)\,\sin^2\xi \,\Big]\;,
\qquad\\[4mm]
  \label{eq:Dparam}
D\,=&\ -v^2 \Big[6 \lambda_1+\lambda_3+\lambda_4  +c_{4 \beta} \Big(2 \lambda_1-\lambda_3-\lambda_4
-\lambda_5\Big) + \lambda_5 + 8 s_{2 \beta} {\rm Re}(\lambda_6)\Big]\;. 
\end{align}
Note that the overall minus sign in the quantity $D$ in~\eqref{eq:Dparam} which emerged after eliminating the 
mass parameters $m^2_{11}$ and $m^2_{22}$ by virtue of the tadpole conditions in~\eqref{Tphi1} and~\eqref{Tphi2}, respectively. In addition, we set $\lambda_1 =\lambda_2$, to showcase the profile of the scalar potential as a function of the CP-odd phase $\xi$ in our numerical examples.

In Figure~\ref{fig:Rxi}, we show the different vacuum topologies that determine the shape of the normalised potential $R_\xi$ [cf.~\eqref{eq:Rxi}] along the phase direction $\xi \in (-\pi,\pi]$, for illustrative choices of $|r_{\rm CP}|$ and $\phi_{\rm CP}$. Moreover, we keep the value of $\tan\beta$ fixed to a specific value, i.e.~${\tan\beta = 2}$ or~$s_{2\beta} = 4/5$. More explicitly, depending on the value of $|r_{\rm CP}|$, the following patterns of CPV may be observed:
\begin{itemize}

\item[{\bf (a)}] {\boldmath $|r_{\rm CP}| >1$}. Then, the 2HDM potential exhibits either CP Conservation (CPC) or violates CP explicitly. This means that besides CPC if $\phi_{\rm CP} = 0$ or $\pi$, only ECPV is possible with a global CPV minimum and a local CPV maximum when $\phi_{\rm CP} \neq 0,\, \pi$, i.e.~the transcendental equation~\eqref{MaxCP} has 2 roots. These features are illustrated in Figure~\ref{fig:Rxi}~(upper left panel).

\item[{\bf (b)}] {\boldmath $1/2< |r_{\rm CP}| <1$}. Then, depending on $\phi_{\rm CP}$, the 2HDM potential will always violate CP in any of the three different forms, SCPV (when $\phi_{\rm CP}=0,\,\pi$), MCPV for small non-zero $\phi_{\rm CP}$ values, and ECPV otherwise. These possibilities were presented in Figure~\ref{fig:Rxi}~(upper right panel). We note that for $|r_{\rm CP}| =1/2$, the critical $\phi_{\rm CP}$ value for transitioning from MCPV to ECPV is: $\phi_{\rm CP}=\pm \pi/4$.

\item[{\bf (c)}] {\boldmath $0< |r_{\rm CP}| <1/2$}. In this case, the 2HDM potential will always violate CP either spontaneously, or
in a mixed spontaneous and explicit manner (MCPV) if $\phi_{\rm CP} \neq 0,\, \pi$.
Moreover, the transcendental equation~\eqref{MaxCP} will have four roots in the primary interval $(-\pi,\pi]$, as shown in Figure~\ref{fig:Rxi}~(lower left panel).

\item[{\bf (d)}] {\boldmath $r_{\rm CP} = 0$}. In this case, the extremal CP-odd constraint~\eqref{MaxCP} will have four solutions: $\xi = 0,\, \pm\pi/2,\, \pi$, if $\lambda_5$ is non-zero and its phase $\varphi_5 = 0$ or $\pi$. Moreover, the expression for the charged Higgs mass in~\eqref{MHp} takes on the simpler form: 
\begin{equation}
   \label{eq:MH+0}
M^2_{H^\pm}/v^2\ =\ -\,\lambda_4\, -\, {\rm Re}(\lambda_5 e^{2i\xi})\,.
\end{equation}
 For $\varphi_5 = \pi$, the extrema would correspond to two CPC minima at $\xi = 0,\, \pi$, as well as to two local CPV maxima at $\xi=\pm\pi/2$, as shown in Figure~\ref{fig:Rxi}~(lower right panel). For $\varphi_5 = 0$, the role of the CPC minima and CPV maxima gets reversed, which could lead to a scenario of SCPV. In both cases, 
 we must have $M^2_{H^\pm}/v^2 > 0$ or $|\lambda_5| > \lambda_4$, if $\lambda_4 > 0$. Also, the breaking of the electroweak symmetry can lead to domain walls. In~addition, if ${\rm Im}\,\lambda_6 \neq 0$, the theory will violate CP explicitly, e.g.~through non-zero scalar-pseudoscalar mass terms~\cite{Pilaftsis:1999qt}. However, the latter source of CPV will not affect the profile of the tree-level 2HDM potential. Its effect will only show up beyond the Born approximation by lifting the degeneracy of the two vacua, thus inducing a scenario of radiative MCPV. 

\end{itemize}

Finally, two remarks are in order here. First, in some of the aforementioned scenarios, there can be instances where the local maximum and the nearby minimum are degenerate, resulting in a horizontal segment of inflection in the scalar potential. Such possibilities require some degree of fine-tuning which we will ignore in our study. 
Second, in all scenarios listed above, only two CPV phases can be at most independent of each other in the general 2HDM. The reason for this is that any third CPV parameter will always be constrained via the CP-odd tadpole condition~\eqref{eq:Ta}, in the weak basis~\eqref{eq:MNbasis} in which no more freedom for any 
Higgs field reparameterisation exists. 

\setcounter{equation}{0}
\section{Continuous Symmetries and Natural Alignment}
\label{sec:CSNAL}

In this section we first briefly review the 2HDM potentials that NHAL is enforced by continuous symmetries. We then identify candidate scenarios that enable sizeable CPV while maintaining agreement with alignment constraints of the SM-like Higgs boson to EW $W^\pm$ and $Z$ bosons. A suitable framework to address this topic is the covariant bilinear field formalism introduced in~\cite{Maniatis:2006fs,Nishi:2006tg,Ivanov:2006yq,Battye:2011jj}. 

To start with, we introduce a $8$-dimensional complex $\bm{\Phi}$-multiplet which represents a vector in the SU(2)$_L\times$Sp(4) field space. The $\bm{\Phi}$-multiplet consists of the scalar iso-doublets~$\Phi_i$ (with $i=1,2$) and their $\mathrm{U(1)}_Y$ hypercharge-conjugate counterparts, $\widetilde{\Phi}_i = i \sigma^{2} \Phi_i^{*}$, i.e.
\bea
   \label{eq:8DPhi}
\bm{\Phi}  \,=\, \begin{pmatrix}
\,\Phi_1 \\ \,\Phi_2 \\ \,\widetilde{\Phi}_1\\  
\,\widetilde{\Phi}_2\:\end{pmatrix}\,, 
\eea
with $\sigma^{1,2,3}$ being the Pauli matrices. The ${\bm{\Phi}}$-multiplet has three essential properties. It transforms covariantly under separate $\mathrm{SU(2)}_L$ and Sp(4) transformations and obeys a Majorana-type constraint~\cite{Battye:2011jj}: 
\begin{equation}
  \label{eq:su2sp4}
\text{(i)}~{\bm{\Phi}'}\, =\,  U_L \, {\bm \Phi}\;,\qquad
\text{(ii)}~{\bm{\Phi}'}\, =\,  U\, {\bm \Phi}\;,\qquad
\text{(iii)}~{\bm{\Phi}}\, =\, C\, {{\bm \Phi}^*}\;, 
\end{equation}
with $U_L\in \mathrm{SU(2)}_L$ and $U\in \text{Sp}(4)$. Moreover,  $C=\sigma^2 \otimes {\bf{1}}_2 \otimes \sigma^2$ is the charge conjugation matrix, which also equips the Sp(4) space with a metric: $U C U^{\sf T} = C$, with~${C=C^{-1}=C^*}$.

Accordingly, the bilinear field vector may be defined as~\cite{Ivanov:2006yq,Pilaftsis:2011ed,Battye:2011jj,Darvishi:2020teg},
\bea
R^A \,\equiv\, {\bm{\Phi}}^{\dagger} \, {\Sigma}^{A}\, {\bm{\Phi}},  \quad A=0, \,1,\,\cdots,\,n(2n-1)-1,
\eea
where the ${\Sigma}^{A}$ matrices
may be expressed in terms of double tensor products as
\bea
{\Sigma}^{A} \, = \, \big(\sigma^0 \otimes t^a_S \otimes \sigma^0, \,\, \sigma^i \otimes t^b_A \otimes \sigma^0 \big),
\eea
with $t^a_S\,\;(t^b_A)\,\in \mathrm{SU}(n)$ as the symmetric (anti-symmetric) generators. Thus, the vector $R^{A} $ explicitly may be written in the following form:
\begin{equation}
R^{A}\: =\: \begin{pmatrix}
\phi_1^{\dagger} \phi_1 +\phi_2^{\dagger} \phi_2 \\ 
\phi_1^{\dagger} \phi_2 +\phi_2^{\dagger} \phi_1 \\ 
-i [\phi_1^{\dagger} \phi_2 -\phi_2^{\dagger} \phi_1] \\
\phi_1^{\dagger} \phi_1 -\phi_2^{\dagger} \phi_2 \\
-[\phi_1^{\mathsf{T}} i \sigma^2 \phi_2 - \phi_2^{\dagger} i \sigma^2 \phi_1^* ]\\
i[\phi_1^{\mathsf{T}} i \sigma^2 \phi_2 + \phi_2^{\dagger} i \sigma^2 \phi_1^* ]
\end{pmatrix}.
\end{equation}
Therefore, by virtue of $R^A$, the potential $\mathcal{V}$ can now be written down in the following quadratic form:
\bea
\mathcal{V} \,=\, -{1\over 2} \,M_A \,  R^A \,+\, {1 \over 4} \, L_{AB} \,  R^{A} \,  R^{B}.
\label{VB}
\eea
In the above, the vector $M_{A}$ and the rank-2 tensor $L_{{A}{B}}$ contain the mass parameters and quartic couplings of the 2HDM potential, respectively. For the 2HDM potential as stated in~\eqref{eq:Vpot}, they assume the following form:
\begin{equation}
   \label{eq:MA}
\setlength{\arraycolsep}{2pt}
\renewcommand{\arraystretch}{0.3}
M_{A}\,=\,\bigg({m_{11}^2 + m_{22}^2\over 2}\,, {\rm Re}(m_{12}^2)\,,
-{\rm Im}~(m_{12}^2)\,, {m_{11}^2 -m_{22}^2\over 2}\bigg) 
\end{equation}
and
{\begin{equation}
   \label{eq:LAB}
\setlength{\arraycolsep}{3pt}
\renewcommand{\arraystretch}{0.6}
L_{{A}{B}}={1\over 2}\begin{pmatrix}
\lambda_1 +\lambda_2 +\lambda_3 & {\rm Re  }(\lambda_6) + {\rm Re }(\lambda_7) &- {\rm Im }(\lambda_6) - {\rm Im }(\lambda_7)  & \lambda_1 - \lambda_2  \\
{\rm Re }(\lambda_6) + {\rm Re }(\lambda_7) & \lambda_4 +{\rm Re }(\lambda_5) & -{\rm Im } (\lambda_5) & {\rm Re }(\lambda_6) - {\rm Re }(\lambda_7) \\
-{\rm Im }(\lambda_6) - {\rm Im }(\lambda_7) & - {\rm Im\,}(\lambda_5) & \lambda_4 -{\rm Re}(\lambda_5) & {\rm Im}(\lambda_7) - {\rm Im }(\lambda_6)  \\
\lambda_1 -\lambda_2 &{\rm Re }(\lambda_6) - {\rm Re}(\lambda_7) & {\rm Im }(\lambda_7) - {\rm Im}(\lambda_6) & \lambda_1 +\lambda_2 -\lambda_3 
\end{pmatrix}.
\end{equation}}

Thus far, several studies have established~\cite{Pilaftsis:2016erj,Darvishi:2019dbh,Darvishi:2020teg,Birch-Sykes:2020btk} that the potential $\mathcal{V}$ of the 2HDM contains 13 ${\mathrm{SU(2)}}_L$-preserving accidental symmetries as subgroups of
the maximal symmetry $\text{SU}(2)_L\otimes \text{SO}(5)$, of which 6 symmetries are $\text{U(1)}_Y$ invariant~\cite{Ivanov:2006yq}.
The symmetry group, $\text{SU}(2)_L\otimes \text{SO}(5)$, which
acts on the 5D bilinear field sub-space $R^I$ (with $I=1,2,3,4,5$), is isomorphic to $\text{SU}(2)_L \otimes \text{Sp}(4)/Z_2$ in the original field space. It plays an instrumental role in classifying accidental symmetries that may occur in the scalar potentials of 2HDM and 2HDM-Effective
Field Theories (2HDMEFT) with higher-order
operators. These classifications were done for the
2HDM in~\cite{Battye:2011jj,Pilaftsis:2011ed,Darvishi:2019dbh} and for the 2HDMEFT framework in~\cite{Birch-Sykes:2020btk}, including higher-order operators of dimension-6 and
dimension-8. 

\begin{table}[ht]
\small
\begin{center}
\begin{tabular}{ |c|c|c|l| } 
 \hline
No. & Generators & Continuous Syms & Parameters\\ \hline 
\hline
$8$  &$T^{0-9}$ & SO(5)&$\ m_{11}^2= m_{22}^2,  \, \lambda_1=\lambda_2=\lambda_3/2.$ \\
\hline 
$7$  &$T^{0,2,5,7,8,9}$ & CP1$\times$O(4)&$\ m_{11}^2= m_{22}^2,  \, \lambda_1=\lambda_2 = \lambda_3/2\,, \lambda_4= -{\rm Re}(\lambda_5).$ \\
$7'$  &$T^{0,1,4,6,8,9}$ & CP2$\times$O(4)$'$&$\ m_{11}^2= m_{22}^2,  \, \lambda_1=\lambda_2 = \lambda_3/2,\, \lambda_4={\rm Re}(\lambda_5).$ \\
$7''$  &$T^{0,3,4,5,6,7}$ & Z$_2\times$O(4)$''$&$\ m_{11}^2= m_{22}^2,  \, \lambda_1=\lambda_2,\, \lambda_3.$ \\
\hline
$6$  &$T^{0,2,5,7,8,9}$ & SO(4) &$m_{11}^2=m_{22}^2,\,  {\rm Im}( m_{12}^{2}),\, \lambda_1= \lambda_2=\lambda_{3}/2,$
\\
& & & $\,\lambda_4= -{\rm Re}(\lambda_5),\,{\rm Im}(\lambda_6)={\rm Im}( \lambda_7).$\\
$6'$  &$T^{0,1,4,6,8,9}$ &  SO(4)$'$ &$m_{11}^2=m_{22}^2,\,  {\rm Re}( m_{12}^2),  \, \lambda_1= \lambda_2=\lambda_{3}/2,$\, 
\\
& & & $\,\lambda_4= {\rm Re}\,(\lambda_5),\, {\rm Re}(\lambda_6)={\rm Re}(\lambda_7).$\\
$6''$  &$T^{0,3,4,5,6,7}$ & SO(4)$''$&$\ m_{11}^2,\, m_{22}^2,  \, \lambda_1,\,\lambda_2,\,\lambda_{3}.$ \\
\hline
$5$&$T^{0,2,4,6}$ & O(2)$\times$ O(3)&$\ m_{11}^2=m_{22}^2, \, \lambda_1=\lambda_2=\lambda_{345}/2,\,\lambda_4={\rm Re}(\lambda_5).$\\
$5'$ &$T^{0,1,5,7}$  &   O(2)$'\,\times$ O(3)&$\ m_{11}^2=m_{22}^2, \, \lambda_1= \lambda_2= \bar{\lambda}_{345}/2,\,\lambda_4=-{\rm Re}(\lambda_5).$\\
$5''$ &$T^{0,3,8,9}$ &    O(2)$''$ $\times$ O(3) & $\ m_{11}^2=m_{22}^2, \, \lambda_1=\lambda_2=\lambda_{3}/2,\,\lambda_4.$\\
\hline
$4$  &$T^{0-3}$ &  O(3)$\times$ O(2)$_Y$ &$\ m_{11}^2=m_{22}^2, \, \lambda_1=\lambda_2= \lambda_{34}/2.$\\
\hline
$3$&$T^{0,5,7}$ &  SO(3)&$m_{11}^2,\, m_{22}^2,\,  {\rm Im}( m_{12}^2),  \, \lambda_1,\, \lambda_2,\, \lambda_{3},\,\lambda_4= -{\rm Re}(\lambda_5),$\\
& & & $\, {\rm Im}(  \lambda_6),\, {\rm Im}(\lambda_7).$
\\
$3'$ &$T^{0,4,6}$ &  SO(3)$'$&$m_{11}^2,\, m_{22}^2,\, {\rm Re}( m_{12}^2),  \, \lambda_1,\, \lambda_2,\, \lambda_{3},\,\lambda_4={\rm Re}(\lambda_5),$\\
& & & $\, {\rm Re}(\lambda_6),\,{\rm Re}(\lambda_7).$
\\
$3''$&$T^{0,8,9}$ & SO(3)$''$&$\ m_{11}^2=m_{22}^2,\,  m_{12}^2,  \, \lambda_1=\lambda_2=\lambda_{3}/2,\,\lambda_4,\, \lambda_5,$ \\
& & & $\,\lambda_6=\lambda_7.$
\\
\hline
$2$&$T^{0,2}$ & CP1$\times$O(2)$\times$O(2)$_Y$ &$ m_{11}^2=m_{22}^2, \,\lambda_1=\lambda_2=\lambda_{345}/2.$ \\
$2'$&$T^{0,1}$ & CP2$\times$O(2)$'\times$O(2)$_Y$ &$ m_{11}^2=m_{22}^2,\,  \lambda_1=\lambda_2=\bar{\lambda}_{345}/2.$ \\
$2''$&$T^{0,3}$ & S$_2\times$O(2)$''\times$O(2)$_Y$ &$ m_{11}^2=m_{22}^2,\,  \lambda_1=\lambda_2,\, \lambda_3,\ \lambda_4.$ \\
\hline
$1$&$T^{0,2}$ & O(2)$\times$O(2)$_Y$&$ m_{11}^2=m_{22}^2, \, {\rm Im}(m_{12}^{2}),  \,\lambda_1=\lambda_2=\lambda_{345}/2,$
\\
& & & $\,{\rm Im}(\lambda_{6})={\rm Im}(\lambda_{7}).$
\\
$1'$ &$T^{0,1}$ & O(2)$'\times$O(2)$_Y$ &  $ m_{11}^2=m_{22}^2, {\rm Re}(m_{12}^{2}),\,  \lambda_1=\lambda_2=\bar{\lambda}_{345}/2$,\, 
\\
& & & ${\rm Re}(\lambda_{6}) ={\rm Re}(\lambda_{7}).$
\\
$1''$&$T^{0,3}$  &  O(2$)''\times$O(2)$_Y$ & $\ m_{11}^2,\,m_{22}^2,  \, \lambda_1,\,\lambda_2,\, \lambda_{3},\,\lambda_4.$\\
\hline
\end{tabular}
\end{center}
\caption{\textit{NHAL and next-to-NHAL symmetries of the 2HDM along with the various SO(5) generators as defined in~\cite{Pilaftsis:2011ed} that reinforce these symmetries. The relations between non-zero parameters associated with these symmetries are given, where we used the abbreviations: $\lambda_{34} \equiv \lambda_3 + \lambda_4$, $\lambda_{345} \equiv \lambda_3 + \lambda_4 + {\rm Re}(\lambda_5)$
and $\bar{\lambda}_{345} \equiv \lambda_3 + \lambda_4 -{\rm Re}(\lambda_5)$. Note that all product groups are subsets of SO(5), so their overall determinant must be evaluated to~1.}}
\label{tab:1}
\end{table}

In Table~\ref{tab:1}, we list all continuous symmetries of the 2HDM potential~\cite{Battye:2011jj,Pilaftsis:2011ed,Darvishi:2019dbh}, where the various choices of the SO(5) generators are displayed according to the conventions of~\cite{Pilaftsis:2011ed}. In the same table, the relationships of the only non-zero mass and quartic-coupling parameters resulting from these symmetries were given as well. Among these symmetries, some of them can accommodate NHAL, in a fashion that it is largely independent\- of the form and size of the soft-symmetry breaking mass parameters added to the potential, or of any specific value that $\tan\beta$ must have~\footnote{Alternatively, SM alignment can be achieved by imposing exact discrete symmetries~\cite{Darvishi:2020teg}, yielding  
\begin{align*}
\tan\beta=1\:,\quad \lambda_1\ =\ \lambda_2, \quad \ \lambda_{3}, \quad \ \lambda_{4}, \quad \ \lambda_{5}, \quad 
\lambda_6\ =\ \lambda_7\ .
\end{align*}
However, such parameter relations mainly lead to an inert Type-I 2HDM in the Higgs basis, for which~NHAL becomes automatic thanks to an unbroken $Z_2$ symmetry.}. Thus, NHAL is obtained when the quartic couplings satisfy the relations~\cite{BhupalDev:2014bir,Darvishi:2020teg},
\begin{align}
  \label{eq:NAcond}
\lambda_1\ =\ \lambda_2\ =\ \frac{\lambda_{345}}{2}, \qquad 
\lambda_6\ =\ \lambda_7\ =\ 0\, . 
\end{align}
with $\lambda_{34} \equiv \lambda_3 + \lambda_4$ and
$\lambda_{345} \equiv \lambda_3 + \lambda_4 + {\rm Re}(\lambda_5)$.
We should stress here that the relations~\eqref{eq:NAcond} must include or take place in a CP-invariant weak basis in which the relative CP-odd phase $\xi$ vanishes, i.e.~when $\xi =0$. Nevertheless, as outlined in Appendix~\ref{app:HAL}, a covariant weak-basis independent condition for SM Higgs alignment can be formulated in terms of a vanishing commutator of two rank-2 Sp(4) tensors as stated in~\eqref{eq:HALcond}.

As mentioned in the introduction, there are three distinct subgroups of NHAL that have been identified so far in the literature,
\begin{eqnarray}
  \label{eq:NHALs}
  \text{\bf Symmetry~8:}&& \text{Sp}(4)\, \simeq\, \text{SO}(5),\ \lambda_1 =\lambda_2 = \frac{\lambda_3}{2}\,,\nonumber\\
  \text{\bf Symmetry~4:}&& \text{SU(2)}_{\rm HF}\times \text{U}(1)_Y\, \simeq\, \text{O}(3)\times \text{O}(2)_Y,\ \lambda_1 = \lambda_2 = \frac{\lambda_{34}}{2}\,,\\
  \text{\bf Symmetry~2:}&& \text{CP}1\times \text{SO}(2)_{\rm HF}\times \text{U}(1)_Y\, \simeq\, \text{CP}1\times \text{O}(2)\times \text{O}(2)_Y,\
  \lambda_1 = \lambda_2 = \frac{\lambda_{345}}{2}\,. \nonumber
\end{eqnarray}
In~\eqref{eq:NHALs}, all $\text{SU}(2)_L$ gauge factors were suppressed, and their isomorphisms in the bilinear field space are given, as well as their listing in Table~\ref{tab:1} of all symmetries that can be relevant to NHAL. While the quartic coupling relations given in~\eqref{eq:NHALs} for Symmetries 8 and 4 remain invariant under $\text{SU}(2)$ reparameterisations of the doublets $\Phi_1$ and $\Phi_2$, this is no longer true for Symmetry 2, for which the relations~\eqref{eq:NAcond} modify in general. In order to put this on a more firm mathematical basis, let us denote the group of reparameterisations with $G_R$, where $G_R = \text{SU}(2)_{\rm HF}$ for the case at hand. Then, the parameter relations obtained from the action of a symmetry group $G$ do not alter, iff the following criterion is satisfied:
\begin{equation}
  \label{eq:Reparam}
  G_R\, \cap\, G\: =\: G_R\; .
\end{equation}
For Symmetry 2, the criterion~\eqref{eq:Reparam} gets violated. In fact, SU(2) field transformations can also change the standard form of CP1. Consequently, any soft-breaking mass parameter introduced in the 2HDM potential must be real~\cite{Pilaftsis:2016erj}. But, as we will analyse in more detail in Section~\ref{sec:SO2HDM}, even in this case, SCPV typically leads to a VEV for $\Phi_2$ with $\xi \neq 0$, thereby triggering deviations from NHAL.

Since CP is a discrete group of central interest to this study, we note that under CP the $R^A$-vector transforms as follows:
\begin{align}{\rm CP} \, R^A\:=\: D_{\text{CP1}} R^A\;,
\end{align}
with $D_{\text{CP1}}=\text{diag} ({\bf{1,1,-1,1,1, -1}}).$
This means that $R^2$ and $R^5$ are odd under these transformations, i.e.~$R^2\to -R^2$ and $R^5\to -R^5$,
whereas all other components of~$R^A$ do not alter.
Therefore, the presence of an odd power of $R^2$'s in the 2HDM potential $\mathcal{V}$ signals candidate scenarios for CPV. Nevertheless, from Table~\ref{tab:1}, we see that all the symmetric cases are CP-conserving, and only a few 
symmetric scenarios exist that could realise 
CPV after the inclusion of soft symmetry-breaking masses in the potential. 

Symmetry groups~$G$ pertinent to Symmetries~$3,5,6,7,8$ and their derivatives that emanate from different field-basis choices under the action of $G_R$ as listed in Table~\ref{tab:1} are collectively referred to as {\em custodial}~\cite{Pilaftsis:2011ed}. They contain generators that do not commute with the generator~$T^0$ of the SM-hyper charge group: ${\text{U}(1)_Y \simeq \text{O}(2)_Y}$. These non-commuting generators can then be used to define the {\em coset space}: $G/\text{O}(2)_Y$, which does not form a {\em quotient group} in general. Hence, these generators are associated with transformations which get violated by the $\text{O}(2)_Y$ gauge coupling $g'$ that occurs in the gauge-kinetic terms of Higgs doublets. On the other hand, Symmetries 4, 2 (2$'$, 2$''$), and 1 (1$'$, 1$''$) are O(2)$_Y$ invariant, and as such, they carry an explicit O(2)$_Y$ factor as displayed in Table~\ref{tab:1}. As a consequence of the above discussion, one must expect that NHAL will be realised by a custodial symmetry group that contains Symmetry~2: CP1$\times$O(2)$\times$O(2)$_Y$. Indeed, from Table~\ref{tab:1}, one may observe that such NHAL symmetry groups, which were not fully accounted for before, exist. They are given by
\begin{eqnarray}
  \label{eq:NHALnew}
\text{\bf Symmetry~7:}&& \text{CP1}\times \text{O}(4)\, \simeq\, \text{CP1}\times\text{Sp}(2)_{\Phi_1\Phi_2}\times \text{Sp}(2)_{\Phi_2\Phi_1}\;,\nonumber\\ 
&&\lambda_1 =\lambda_2 = \frac{\lambda_{3}}{2}\,,\ \lambda_4 = -\text{Re}(\lambda_5)\,,\nonumber\\
\text{\bf Symmetry~5:}&& \text{O}(2)\times \text{O}(3)\, \simeq\, \text{SO}(2)_{\rm HF}\times \text{Sp}(2)_{\Phi_1+\Phi_2},\nonumber\\ 
&& \lambda_1 =\lambda_2 = \frac{\lambda_{345}}{2}\,,\ \lambda_4 = \text{Re}(\lambda_5) \;. 
\end{eqnarray}
We note the embedding: CP1$\times$O(2)$_Y\subset\text{Sp}(2)_{\Phi_1+\Phi_2}$. The latter as well as the custodial subgroups $\text{Sp}(2)_{\Phi_1\Phi_2},\,\text{Sp}(2)_{\Phi_2\Phi_1}  \subset \text{Sp}(4)$ are all defined in~\cite{Darvishi:2019dbh}.
Likewise, it is interesting to notice that Symmetry 2 goes to the higher Symmetry~5 in the limit: $\text{Re}(\lambda_5) \to \lambda_4$, and Symmetry 7 to 8, when $\lambda_4 \to 0$. 

We should clarify here that Symmetry 7 was classified before in~\cite{Pilaftsis:2011ed} as a subgroup of the symplectic group Sp(4), which is one of the three primary realizations for NHAL. Recently, a set of parameter relations similar to Symmetry 7 leading to NHAL was also observed in~\cite{Aiko:2020atr}, within 
the context of some twisted custodial group: $\text{SU}(2)_R\times \text{SU}(2)_L$ that includes the $\text{SU}(2)_L$ gauge group. In this respect, we should comment here that this twisted group is only a subgroup of the group: $\text{CP1}\times\text{Sp}(2)_{\Phi_1\Phi_2}\times \text{Sp}(2)_{\Phi_2\Phi_1}\times \text{SU}(2)_L$,
as~specified in~\eqref{eq:NHALnew}.
Moreover, to the best of our knowledge, Symmetry~5 given in~\eqref{eq:NHALnew}, which was tabulated in~\cite{Battye:2011jj,Pilaftsis:2011ed}, has not been analysed before in connection with NHAL. 

\begin{table}[t]
\small
\begin{center}
\begin{tabular}{ |c|| c| c| c| c| } 
 \hline
No. & Syms with NHAL & No. & Syms breaking of NHAL & Types of CPV after SB\\ 
\hline \hline
8 & SO(5)& \multicolumn{3}{c|}{---} \\
\hline
7 & CP1$\times$O(4) & 6 & SO(4)& ECPV/MCPV
\\
7$'$  &  CP2$\times$O(4)$'$  & 6$'$ & SO(4)$'$& SCPV\\
\hline
5 &  O(2$)\times$O(3)  & 6$'$ & SO(4)$'$& SCPV
\\
5$'$  & O(2$)'\times$O(3) & 6 & SO(4)& ECPV/MCPV\\
\hline
4 & O(3)$\times$O(2)$_Y$ & \multicolumn{3}{c|}{---} \\
\hline
2 & CP1$\times$O(2)$\times$O(2)$_Y$ & 1 & O(2)$\times$O(2)$_Y$ & ECPV/MCPV\\
2$'$  & CP2$\times$O(2)$'\times$O(2)$_Y$  & 1$'$  & O(2)$'\times$O(2)$_Y$ & SCPV\\
\hline
\end{tabular}
\end{center}
\caption{\textit{The continuous symmetries resulting from the CP breaking of NHAL symmetries are presented. The final column indicates the types of CPV realised after the introduction of soft symmetry-breaking mass terms.}}
\label{tab:2}
\end{table}

In Table~\ref{tab:2}, we have explored whether the five NHAL symmetries given in~\eqref{eq:NHALs} and~\eqref{eq:NHALnew} (and those descending from different field-basis choices) can lead to physical~CPV or not. However, we must emphasise that not all NHAL symmetries can lead to CPV, after introducing soft symmetry-breaking masses, but only those associated with CP1 and CP2 symmetries. For instance, Symmetries 8 and 4 cannot source CPV, unless an explicit hard-breaking of symmetries is considered. Instead, as shown in the fourth column of Table~\ref{tab:2}, Symmetries $2$ and $2'$ can
break to the lower symmetries $1$ and $1'$, after specific operators (breaking softly or explicitly $2$ and $2'$) are added to the potential. Interestingly enough, Symmetries~$6,\,6'$ will transition to the lower symmetries~$5',\,5$, when the symmetry-breaking parameters of the latter are switched off from the potential. This paradox can be resolved by noting that all $6$-type symmetries are based on $\text{SO}(4) \simeq \text{SO}(3)\times \text{SO}(3)$ containing two custodial factors that produce no new restrictions on a U(1)$_Y$-invariant 2HDM potential. They could only produce non-trivial constraints on the new theoretical parameters present in a hypothetical U(1)$_Y$-violating 2HDM potential${}$~\cite{Battye:2011jj}.  

The types of CPV that can be realised by adding now all possible soft symmetry-breaking mass terms have been catalogued in the last column of Table~\ref{tab:2}. Our approach to maximising CPV should be characterised as natural according to 't Hooft's naturalness criterion, since an enhanced symmetry gets realised once all CPV terms are switched off.  

In the next section, our attention will be directed towards approximate symmetric 2HDM scenarios that allow for observable CPV as given by the fourth column of Table~\ref{tab:2}. In particular, we will investigate if a viable parameter space exists that 
can realise sizeable CPV, while maintaining agreement with limits from the non-observation of an electron EDM and from
SM alignment and other LHC data.

\setcounter{equation}{0}
\section{\boldmath O(2)$\times$O(2)$_Y$ and SO(4) Symmetric 2HDMs}
\label{sec:SO2HDM}

In the previous section, we have found that in addition to CP1$\times$SO(2)$\times$O(2)$_Y$, there exist two new custodial NHAL symmetries based on the product groups: $\text{CP1}\times\text{O}(4)$ and $\text{O}(2)\times \text{O}(3)$ [cf.~\eqref{eq:NHALnew}]. From Table~\ref{tab:2}, we see
that CP1$\times$SO(2) and CP1$\times$O(4) break into the symmetry groups, O(2)$\times$O(2$)_Y$ and SO(4), respectively. As we will explicitly demonstrate in this section, these two groups are the ones that enable one to build minimal scenarios of soft and explicit CP breaking while maximising CPV in the scalar potential.

Let us first consider the parameter space of O(2$)\times$O(2)$_Y$-symmetric 2HDM. In the bilinear field formalism,
this model is invariant under the action of the $T^0$ and $T^1$ generators, which amounts to a 2D rotation about the $R^0$-$R^2$ plane. In the original scalar field space, this can be equivalently described by a transformation of the form:
\begin{align}
\Phi_+\, =\, \frac{1}{\sqrt{2}}\,\big(\Phi_1 + i \Phi_2\big)\ &\to\ 
\Phi'_+\, =\, \frac{e^{i \alpha}}{\sqrt{2}}\, \big(\Phi_1 +  i\Phi_2 \big)\,,
\nonumber \\
\Phi_-\, =\, \frac{1}{\sqrt{2}}\,\big(\Phi_1 - i \Phi_2\big)\ &\to\
\Phi'_-\, =\, \frac{e^{-i \alpha}}{\sqrt{2}} \big(\Phi_1 -  i\Phi_2 \big)\,. 
\end{align}
Therefore, in an O(2)$\times$O(2)$_Y$-invariant 2HDM, 
the potential parameters obey the following relations:
\begin{eqnarray}
{\rm O}(2)\times {\rm O}(2)_Y:  \quad m_{11}^2 = m_{22}^2,\ \ {\rm Im}(m_{12}^{2}),\  \ \lambda_1 = \lambda_2 = \frac{\lambda_{345}}{2},\ \ {\rm Im}(\lambda_{6}) = {\rm Im}(\lambda_{7})\,.
\end{eqnarray}
Observe that the naively CP-odd parameter ${\rm Im}(m_{12}^{2})$ is independent in this model. Nevertheless, the O(2)$\times$O(2)$_Y$-symmetric potential is CP conserving, which can be broken softly by assuming non-zero mass terms, such as ${\rm Re}(m_{12}^2)$ and $m_{11}^2 - m_{22}^2 \neq 0$. As a consequence of the addition of these two terms, a non-removable CP-odd phase will appear in the 2HDM potential. 

Let us now turn our attention to the SO(4)-symmetric 2HDM. 
The parameter space of this model is constrained by the action of
generators $T^{0,2,5,7,8,9}$ which correspond to Symmetry~6 in 
Table~\ref{tab:1}. Therefore, the parameters of this SO(4)-symmetric model satisfy the relations,
\begin{equation}
\mathrm{SO}(4):\quad m_{11}^2 = m_{22}^2, \quad  \mathrm{Im}(m_{12}^{2}), \quad \lambda_1 = \lambda_2 = \frac{\lambda_{3}}{2},\quad \mathrm{Re}(\lambda_{5}) = -\lambda_4,\quad  \mathrm{Im}(\lambda_{6}) = \mathrm{Im}(\lambda_{7})\,.
\end{equation}
As before, one could introduce soft symmetry-breaking mass terms 
in the SO(4)-symmetric potential in order to allow for CPV.
In this context, the breaking pattern: SO(4) $\to$ O(2)$\times$O(3) $\to$  SO(3), takes place. These breakings produce small departures from NHAL, while violating the CP symmetry of the 2HDM potential. If we now go to a weak basis where $\lambda_6 = \lambda_7$, we will see that all three types of CPV can be realised.

\begin{table}[t!]
\centering
\resizebox{1\textwidth}{!}{
\begin{tabular}{|c|c|c|c|c|c|c|c|c|c|}
\hline
Symmetries & CPV types & $\tan\beta$ & $\xi$ & $\phi_{12}$ & $M_{H_1}$ & $M_{H_2}$ & $M_{H_3}$ & $M_{H^\pm}$ & $g_{H_1VV}$ \\
\hline
\multirow{2}{*}{O(2)$\times$O(2)$_Y$} & ECPV & 1.608 & 0.002 & -0.05  & 125.10 & 682.99 & 787.67 & 640.18 & 0.998 \\
& MCPV & 0.804 & -1.93& 0.27 & 125.21 & 225.11 & 270.50 & 380.54 & 0.989 \\
O(2)$'\times$O(2)$_Y$& SCPV & 0.71 & 0.67 & 0  & 125.52 & 275.23 & 448.01 & 551.38 & 0.957 \\
\hline
\multirow{2}{*}{SO(4)} & ECPV & 0.827 & 0.002 & -0.02 & 125.28 & 506.93 & 570.58 & 570.59 & 0.995 \\
& MCPV & 0.740 & -1.25 & 0.44  & 125.40 & 408.49 & 433.98 & 433.99 & 0.984 \\
SO(4)$'$& SCPV & 0.742 & -2.42 & 0 & 125.68 & 216.63 & 265 & 442.21 & 0.948 \\
\hline
\end{tabular}}
 \caption{\textit{Benchmark scenarios to be used in our analysis shown in Figures~\ref{ECPV-AL}, \ref{MCPV-AL} and \ref{SCPV-AL}. All angles are given in radians and all masses in units of~GeV.}}
\label{tab:BMs}
\end{table}

In order to assess the allowed size of CPV, we will perform a scan over the model parameters, ensuring that they also align with experimental constraints on gauge couplings.
The key model parameters for our scan are
\begin{align}
\tan\beta\,,\ M_H^\pm\, ,\ \lambda_{3}\, ,\ \lambda_{4}\,,\ \xi\,,\ \phi_{12}\;.
\end{align}
Furthermore, the following restrictions on the parameters were imposed:
\begin{equation}
M_H^\pm \in [250, 800]\, {\rm GeV}\,,\quad \beta \in \left(0, \frac{\pi}{2}\right), \quad \{\xi,\phi_{12}\} \in (-\pi, \pi]\;,
\end{equation}
along with $M_{H_1} = 125.46 \pm 0.35$ GeV. As for the quartic couplings $\lambda_{3,4}$, these are only constrained to lie in the perturbative regime, i.e.~$\lambda_{3,4} < 4\pi$. For the SO(4)-symmetric scenarios, the relationship $\lambda_{3} = 2\lambda_{1}$ holds. Consequently, after a careful parameter scan, the benchmark models of interest to us are summarized in Table~\ref{tab:BMs}.

In addition, the stringent experimental limit on the electron EDM~\cite{Andreev:2018ayy},
\begin{align}
   \label{eq:exp-edm}
|d_e|\, <\, 1.1 \times 10^{-29} \ e\cdot {\rm cm}
\end{align}
at the 90\% confidence level, puts severe constraints on any new source of CPV including those that come from the 2HDM potential. The effective Lagrangian that describes the interaction of a non-zero EDM of a fermion with the electromagnetic field reads
\begin{align}
\mathcal{L}_{\rm EDM}\, =\, -{i\over 2} d_f \bar{\psi} \sigma^{\mu \nu} \gamma^5 \psi F_{\mu \nu}\,.
\end{align}
In the SM, the electron EDM is predicted to be several orders below the 
current experimental sensitivity. For this reason, any observation of a non-zero electron EDM will be a signal of new physics, which could be sourced from 
an extended CPV scalar sector of the 2HDM. In our analysis, we should not only satisfy the EDM upper limit~\eqref{eq:exp-edm}, but also other constraints that result from LHC data.

\begin{table}[t]
\centering
\resizebox{1\textwidth}{!}{%
\begin{tabular}{|c|c|c|c|c|c|c|c|c|c|}
\hline
\multirow{2}{*}{Symmetries} & \multirow{2}{*}{CPV types} & \multicolumn{7}{c|}{The ratio of EDM contributions over the total EDM} & \multirow{2}{*}{EDM [$e\cdot {\rm cm}$]} \\
\cline{3-9}
& & \eqref{A.1} & \eqref{A.2} & \eqref{A.3} & \eqref{A.4} & \eqref{A.5} & \eqref{A.6} & \eqref{A.7} & \\
\hline
\multirow{2}{*}{O(2)$\times$O(2)$_Y$} & ECPV & 5.581 & -0.072 & -4.711 & -0.308 & 0.0185 & 0.312 & 0.180 & $9.51 \times 10^{-30}$ \\
& MCPV & 2.008 & -0.026 & -0.942 & -0.065 & 0.0015 & 0.028 & -0.005 & $7.67 \times 10^{-30}$ \\
O(2)$'\times$O(2)$_Y$& SCPV & 2.003 & -0.026 & -0.866 & -0.079 & 0.0002 & 0.004& -0.037 & $3.86 \times 10^{-30}$ \\
\hline
\multirow{2}{*}{SO(4)}  & ECPV & 1.851 & -0.021 & -0.832 & -0.055 & 0.0005 & 0.009& 0.048 & $2.87 \times 10^{-30}$ \\
& MCPV & 1.995 & -0.021 & -0.887 & -0.077 & -0.0013 & -0.023 & 0.016 & $1.07 \times 10^{-30}$ \\
SO(4)$'$& SCPV & -1.211 & 0.014 & 1.875 & 0.172 & 0.0002 & 0.004 & 0.146 & $3.55 \times 10^{-30}$ \\
\hline
\end{tabular}}
\caption{\textit{Individual contributions to the electron EDM that originate from the seven two-loop graphs displayed in~\eqref{A.1}--\eqref{A.7}.}}
\label{tab:EDMcontrib}
\end{table}

In the Type-II 2HDM, the dominant contribution to the electron EDM originates at two loops from the so-called Barr--Zee mechanism~\cite{Barr:1990vd,Pilaftsis:1999qt,Abe:2013qla,Darvishi:2022wnd}. Instead, the one-loop contributions to $d_e$ are rather suppressed. The analytic expressions of the two-loop contributions to the electron EDM are given in Appendix~\ref{app:eEDM} [cf.~\eqref{A.1}--\eqref{A.7}]. In Table~\ref{tab:EDMcontrib},
we exhibit the individual contributions to the electron EDM for the benchmark scenarios given in~Table~\ref{tab:BMs} which are all compatible with the experimental constraint~\eqref{eq:exp-edm}.

From Table~\ref{tab:EDMcontrib}, we must also notice that the different EDM contributions do not separately exceed the electron EDM limit for most of the benchmark models. An exception constitutes the O(2)$\times$O(2)$_Y$-symmetric 2HDM with ECPV, whose viability would require mild cancellations to take place in less than one part in ten, among the EDM effects evaluated in~\eqref{A.1},~\eqref{A.2} and~\eqref{A.3}. It is important to remark here that the current electron EDM limit as deduced from ThO~\cite{Andreev:2018ayy} gives constraints on the size of new CPV phases that are stronger by almost two orders of magnitude than other EDM bounds from neutron and Mercury~\cite{Baker:2006ts,Griffith_2009,Pendlebury:2015lrz}. As a consequence, no further arrangement for cancellations would be needed to eliminate potential contributions coming from the chromo-EDM and the three-gluon Weinberg operators, as was often done in the past within the context of supersymmetric theories~\cite{Ellis:2008zy}.

\begin{figure}[t]
\centering
\includegraphics[width=1.01\textwidth]{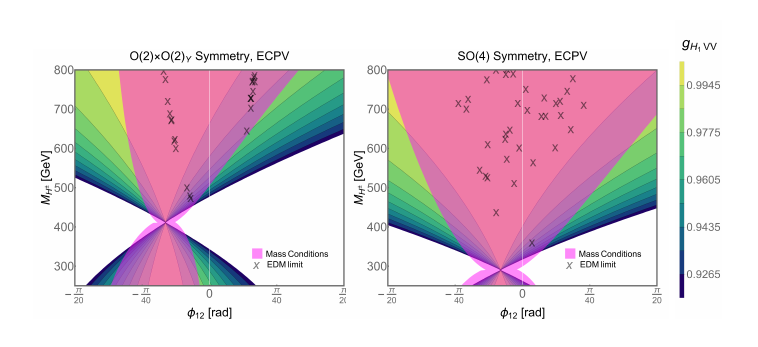}
\includegraphics[width=1.01\textwidth]{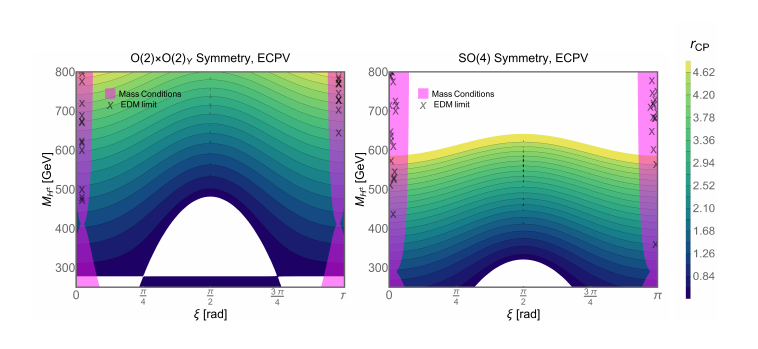}
\caption{\textit{The gauge boson coupling $g_{H_{1}VV}$ and $r_{\rm CP}$ in the ($M_H^\pm, \xi$)-plane are displayed for different types of ECPV in O(2$)\times$O(2)$_Y$- and SO(4)-symmetric 2HDMs. The cross symbols ~`$\times$' showcase benchmark points consistent with the electron's EDM limit in Eq.~\eqref{eq:exp-edm}. }}
\label{ECPV-AL} 
\end{figure}

{\allowdisplaybreaks
\begin{figure}[t]
\centering
\includegraphics[width=1.01\textwidth]{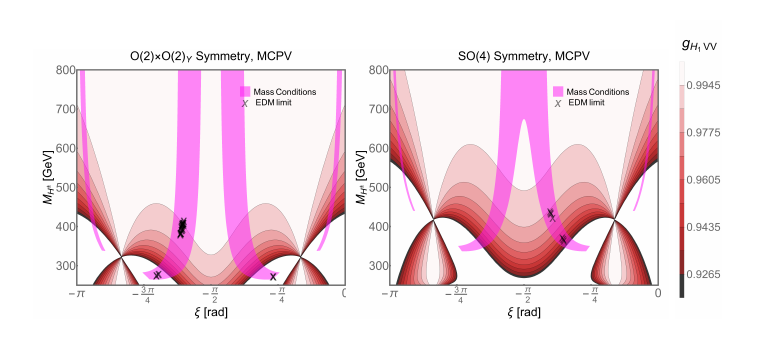}
\\
\includegraphics[width=1.01\textwidth]{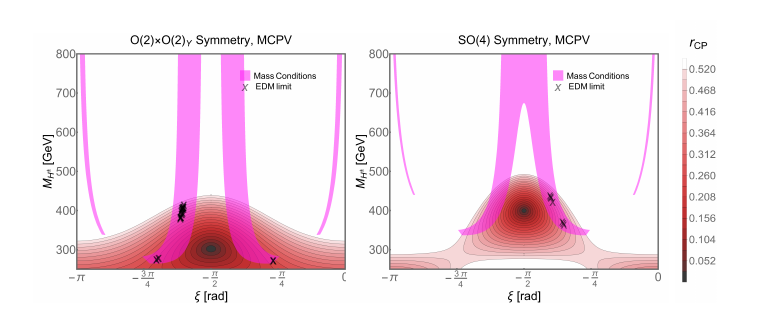}
\caption{\textit{The same as in Figure~\ref{ECPV-AL}, but for MCPV. }}
\label{MCPV-AL} 
\end{figure}}

{\allowdisplaybreaks
\begin{figure}[t]
\centering
\includegraphics[width=1.01\textwidth]{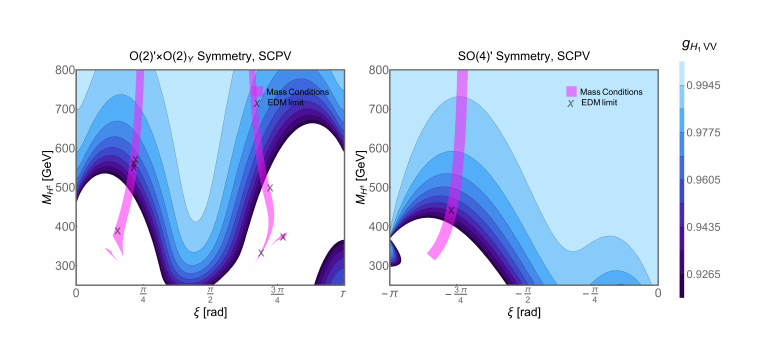}
\includegraphics[width=1.01\textwidth]{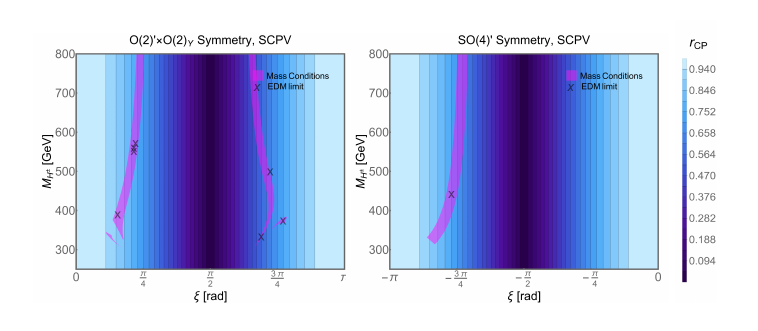}
\caption{\textit{The same as in Figure~\ref{ECPV-AL}, but for SCPV in O(2)$'\times$O(2)$_Y$- and SO(4)$'$-symmetric 2HDMs.}}
\label{SCPV-AL} 
\end{figure}}

In Figures \ref{ECPV-AL}, \ref{MCPV-AL} and~\ref{SCPV-AL}, we show the predicted values for the SM-normalised coupling $g_{H_1VV}$ of the lightest $H_1$-boson to the gauge bosons $V=W^\pm,Z$ (upper panels) and the parameter~$|r_{\rm CP}|$ (lower panels) in the O(2$)\times$O(2)$_Y$ (left panels) and SO(4) (right panels) symmetric 2HDMs for three different types of CP-violating scenarios: ECPV, MCPV, and SCPV. These three types of CPV scenarios are depicted in green, red, and blue colours, respectively. In our numerical estimates, the CP-odd phases $\phi_{12}$ and $\xi$ are varied, so as to obey the two mass conditions: (i)~$M_{H_1} \approx 125$~GeV and (ii)~$M_{H_3} > M_{H_2}$. In these plots,  we provide a set of benchmark points shown with cross symbols~`$\times$' that are in agreement with the electron's EDM limit in Eq.~\eqref{eq:exp-edm}. It is important to note that this does not validate or exclude the rest of the region.

In a 2HDM scenario with ECPV, the phase $\phi_{12}$ plays a crucial role, as shown in the upper panels of Figure \ref{ECPV-AL}, where the predicted values of the SM-normalised $H_1VV$-coupling, $g_{H_1VV}$, are projected onto the $(\phi_{12},M_{H^\pm})$-plane. By analogy, Figures~\ref{MCPV-AL} and~\ref{SCPV-AL} offer similar projections for the SCPV and MCPV scenarios, respectively. They illustrate how the size of $g_{H_1VV}$ gets distributed on the $(\xi ,M_{H^\pm})$-plane. 
In all these figures, the violet area highlights acceptable regions consistent with Higgs mass $M_{H_1}=125.46\pm 0.35$ GeV and $M_{H_3}>M_{H_2}>200$ GeV, while the regions that are excluded from the analysis are shown without colour. As can be seen from Figures~\ref{ECPV-AL}, \ref{MCPV-AL} and~\ref{SCPV-AL}, a set of benchmark points can simultaneously satisfy all the constraints.

Further insight can be gleaned from the lower panels of Figures~\ref{ECPV-AL},~\ref{MCPV-AL} and~\ref{SCPV-AL}. For each parameter point of the top panels, there exists a corresponding value for $|r_{\rm CP}|$. Observe that in the lower right frame of Figure~\ref{ECPV-AL}, EDM values located in the white space correspond to $|r_{\rm CP}| > 5$. This interpretation and the respective $|r_{\rm CP}|$ values are in line with the discussion presented in Section \ref{sec:CPV2HDM}. Note that scenarios, with $\lambda_4 = (-){\rm Re}(\lambda_5)$, predict a mass degeneracy between the charged Higgs bosons and the heavy neutral Higgs states. However, this mass degeneracy gets lifted if the phase~$\xi$ deviates significantly from zero. We may reiterate here that the relevant benchmark scenarios we have been analysing are given in~Table~\ref{tab:BMs}, including an SO(4)$'$-symmetric 2HDM in which one has $\lambda_4 = {\rm Re}(\lambda_5)$
according to Table~\ref{tab:1}.

In a CPV framework of the 2HDM, there exist additional contributions to $ZH_1$ production channel, primarily through the resonant production and decay of the heavy $H_{2,3}$ bosons~\cite{Keus:2015hva,Bian:2016awe}. At the LHC, the dominant production mechanism of the $H_{2,3}$ bosons is the gluon-gluon fusion process. Furthermore, non-SM modifications to the $H_1 ZZ$ coupling may induce relevant deviations in the $ZH_1$ production cross-section. Therefore, it is crucial to assess the magnitude of these deviations, while being in good agreement with established SM Higgs properties.

\begin{figure}[t]
\begin{center}
\includegraphics[width=0.8\textwidth]{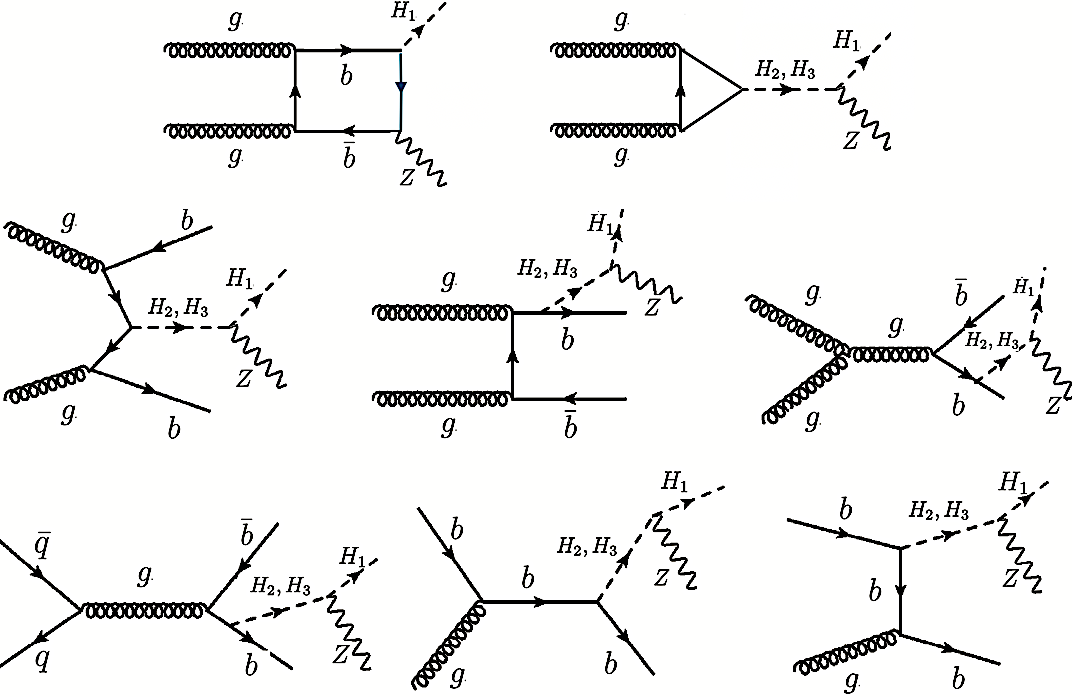}
\caption{\textit{The leading-order Feynman diagrams for the production of heavy states $H_2, H_3$ in association with $b$-type quarks, followed by their decay to $ZH_1$. For SM contributions heavy states $H_2, H_3$ can be substituted with $Z$ boson.}}
\label{ch-ZH}
\end{center}
\end{figure}

Figure \ref{ch-ZH} displays a representative set of leading-order Feynman diagrams for the production of the heavy states $H_2, H_3$ in association with $b$-type quarks, along with their subsequent decay into $ZH_1$. Diagrams obtained by switching the initial state gluons or by emitting the Higgs from an $\bar{b}$-type quark are omitted for brevity. These heavy states possess loop-induced couplings to two gluons, which are generated by their tree-level $H_it\bar{t}$ couplings. As a result, they are predominantly produced via gluon-gluon fusion at the~LHC. Alternatively, if the bottom-quark Yukawa coupling is sufficiently large, another promising production mechanism becomes significant that involves $H_2, H_3$ production in association with two bottom quarks. The resulting signals at the LHC will mainly stem from the decays of $H_{2,3}$ bosons to the SM-like Higgs boson $H_1$ in association with a $Z$ boson, i.e.~$pp \to H_{2,3} \to ZH_1$.

\begin{figure}[t]
\centering
\includegraphics[width=0.7\textwidth]{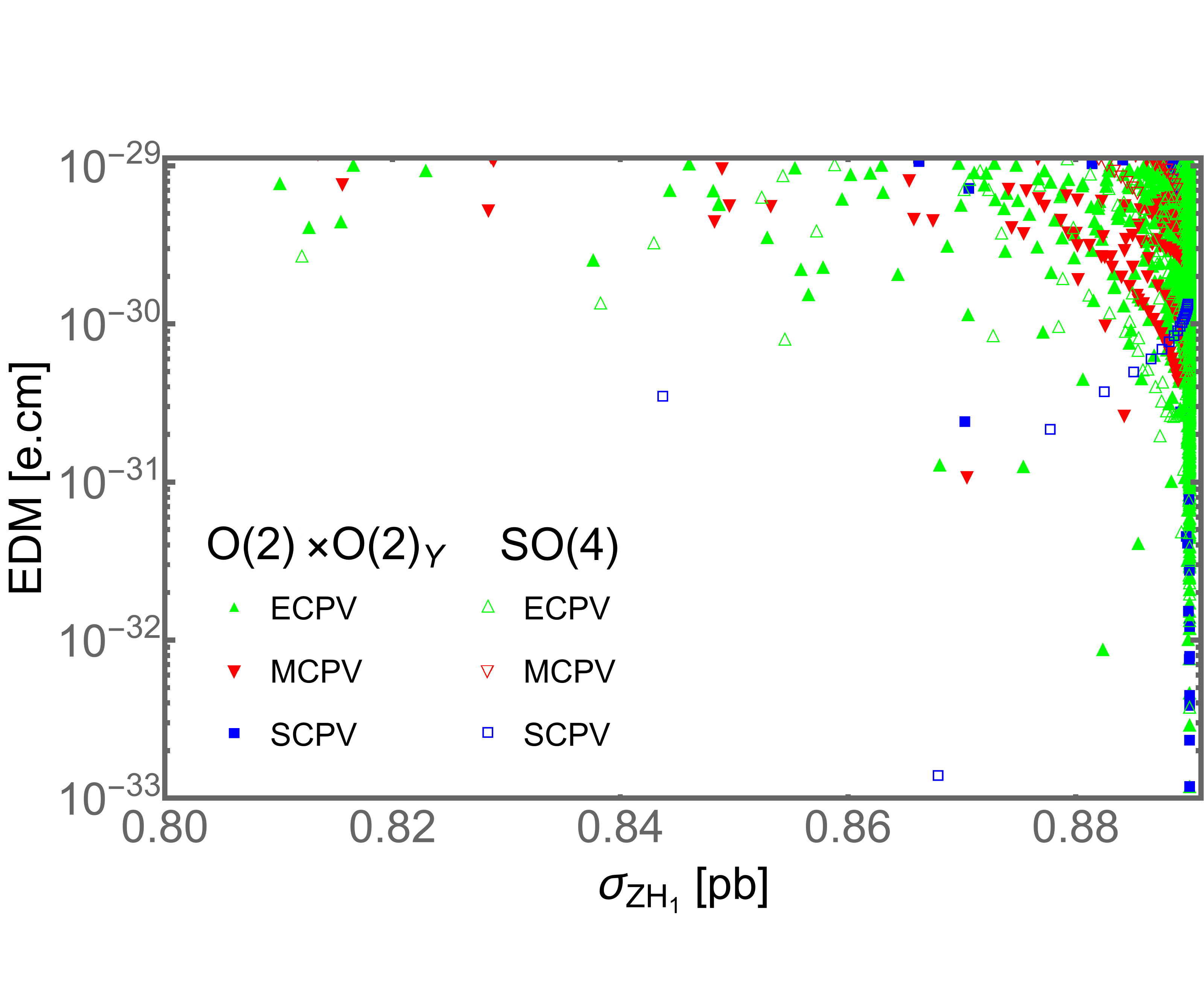}
\caption{\textit{The $ZH_1$ production cross-section is displayed for O(2$)\times$O(2)$_Y$- and SO(4$)$-symmetric models compatible with the current electron EDM limit given in~\eqref{eq:exp-edm}.}}
\label{fig:ZH}
\end{figure}

The production cross section in proton-proton collisions, $\sigma(pp \to H_1Z)$, may be approximated as
\begin{equation}
  \label{eq:sigmapp}
\sigma(pp \to H_1Z)\, \approx\, \sigma(pp \to H_1(M_{H_1})) \times g_{H_1ZZ} + \sum_{i=2,3} \sigma(pp \to H_i(M_{H_i})) \times g_{H_iH_1Z}\;,
\end{equation} 
where the couplings $g_{H_i H_1 Z}$ can be replaced with the SM-normalised $g_{H_jZZ}$ couplings as follows:
\begin{equation}
g_{H_i H_j Z}\, =\, \epsilon_{ijk}\,g_{H_k ZZ}\,,
\end{equation}
with $i,j=1,2,3$. Equation~\eqref{eq:sigmapp} includes both the direct $H_1$ production and the associated $H_1Z$ production via~$H_{2,3}$ decays. We should note that in 2HDMs with exact NHAL, the couplings of the heavy $H_{2}$ and $H_{3}$ bosons to $H_1Z$ vanish at tree level, and so only the first top left Feynman graph in Figure~\ref{ch-ZH} will contribute significantly~\cite{BhupalDev:2014bir}. 
Figure~\ref{fig:ZH} projects the cross-section for 
$ZH_1$ production for the O(2$)\times$O(2)$_Y$- and SO(4)-symmetric 2HDMs. The colour code of this plot is the same as in Figure~\ref{fig:Rxi}. These results are below the current electron EDM upper limit given in~\eqref{eq:exp-edm}. 
Finally, we should remark that approximately 98\% of the total production cross section as estimated in~\eqref{eq:sigmapp} arises from the top left one-loop graph in Figure~\ref{ch-ZH} and only 2\% from the remaining contributions that involve the exchanges of the heavy $H_{2,3}$ bosons.

\section{Conclusions}\label{sec:Concl}

We have presented a detailed study of the Two Higgs Doublet 
Model~(2HDM) by paying\- special attention to the vacuum topology of its potential and its connection to CP Violation (CPV). 
We have rigorously identified that in general three types of CPV exist: (i) Spontaneous CPV, (ii) Explicit CPV and (iii) Mixed Spontaneous and Explicit CPV (MCPV). In all these three different scenarios, only two CPV phases can remain independent, since any third CPV parameter will be constrained by the CP-odd tadpole condition given in~\eqref{eq:Ta}.

The vacuum topology of a general 2HDM potential is usually determined by the action of some accidental symmetry on the potential, as well as the size of any explicit or spontaneous breaking parameter of these symmetries, like the CP-odd phases $\xi$ and $\phi_{12}$. In this respect, an important finding of the present work was to exemplify the pivotal role of the complex parameter~$r_{\rm CP}$ defined in~\eqref{eq:rCP}, which enables one to accurately differentiate the three different realisations of CPV mentioned above. The value of~$r_{\rm CP}$, namely~its magnitude and CP-odd phase~$\phi_{\rm CP}$, can be used to characterise the CPV-type of the 2HDM. Although it is well known that for $|r_{\rm CP}| > 1$ only an ECPV scenario will be feasible, 2HDM settings with $|r_{\rm CP}| < 1$ can exhibit a much richer structure than that was considered before when sources of explicit CPV are allowed. Specifically, we have found that depending on the value of~$\phi_{\rm CP}$, all types of SCPV, ECPV and MCPV, can individually take place if $1/2 < |r_{\rm CP}| < 1$.
On the other hand, if $0 < |r_{\rm CP}| < 1/2$, we can only have SCPV and MCPV, so there will always be a degree of SCPV present in this case. Finally, a 2HDM scenario 
with $r_{\rm CP}$ tuned to zero can have a more complex topology 
and so the inclusion of higher-order effects on the potential beyond the Born approximation will be necessary. The profiles of the different CPV-types of 2HDM potentials have been illustrated in Figure~\ref{fig:Rxi}, in terms of $R_{\xi}$ defined in~\eqref{eq:Rxi}, as functions of the phase~$\xi$.

An important phenomenological constraint on any theory of new physics is the required\- alignment of the coupling strengths of the observed 125-GeV scalar resonance to the $W^\pm$ and $Z$ gauge bosons with strengths as predicted by the SM. In the 2HDM, such an SM-Higgs alignment can be naturally achieved by invoking on certain accidental symmetries~\cite{BhupalDev:2014bir}. However, their imposition leads unavoidably to a CP-conserving 2HDM. Here, we have revisited these accidental symmetries of NHAL by employing the bilinear covariant formalism, with the aim of maximising CPV through minimal soft and explicit breakings into lower symmetries. To this end, we first recovered the two known NHAL symmetries that can act on a general CP-violating 2HDM potential, i.e.~Sp(4) and SU(2)$_{\rm HF}$ in the original field basis. But when CP is imposed on a general 2HDM potential thanks to a CP1 or CP2 symmetry, we have then found that besides CP1$\times$O(2)$\times$O(2)$_Y$, there exist two new custodial NHAL symmetries that make use of the product groups: $\text{CP1}\times\text{O}(4)$ and $\text{O}(2)\times \text{O}(3)$ [cf.~\eqref{eq:NHALnew}]. In fact, the groups, CP1$\times$O(2)$\times$O(2)$_Y$ and CP1$\times$O(4), break into the key symmetry groups, O(2)$\times$O(2$)_Y$ and SO(4), respectively. The latter two groups are the ones that enable one to construct minimal scenarios of soft and explicit CP breaking as well as maximise CPV in the scalar potential.

We have derived upper limits from the non-observation of an electron EDM on key CPV parameters that quantify the degree of SM misalignment in representative scenarios based on the two symmetry groups mentioned above. In addition, we have delineated the CPV parameter space of such 2HDM scenarios with approximate NHAL that could be probed at the LHC. In Table~\ref{tab:BMs}, we have presented CPV scenarios that conform with experiment by highlighting their compatibility with the electron EDM limit~(see Table~\ref{tab:EDMcontrib}) and the predicted $ZH_1$ production rate at the LHC as shown in~Figure~\ref{fig:ZH}.

The present work offers new directions for future research by providing strategies on how to naturally maximise CPV beyond the framework of 2HDMs with approximate NHAL that we have studied here. Likewise, it would be interesting to investigate whether successful scenarios of electroweak baryogenesis scenarios can be realised without resorting to fine-tuning, within the context of the naturally aligned 2HDMs similar to those presented in this paper.  

\subsection*{Acknowledgements} 

The works of ND and AP are supported by STFC under the grant numbers ST/T006749/$1$ and ST/X00077X/$1$, respectively.
The work of J.H.Y. is supported by the National Science Foundation of China under Grants No. 12347105, and No. 12375099, and National Key Research and Development Program of China Grant No. 2020YFC2201501, and No. 2021YFA0718304.

\vfill\eject
\appendix  

\renewcommand{\theequation}{\Alph{section}.\arabic{equation}}

\setcounter{equation}{0}
\section{Higgs Masses and Couplings}\label{app:2.2}

CPV in the Higgs potential gives rise to mixing between CP-even and CP-odd scalar fields. To account for this mixing,
one has to consider a $(3\times 3)$-dimensional mass matrix for  the neutral Higgs bosons.
Henceforth, physical neutral states of definite mass ($H_1, H_2, H_3$) can be obtained by an orthogonal transformation upon the weak basis $(\phi_1\,,\phi_2\,,a)$ as follows:
\begin{align}
\begin{pmatrix}
H_1 \\
H_2 \\
H_3 \\
\end{pmatrix}
=
\mathcal{R}
\begin{pmatrix}
\phi_1 \\
\phi_2 \\
a
\end{pmatrix}.
\end{align}
In the above, $\mathcal{R}$ is a $3\times 3$ orthogonal matrix that may conveniently be written as
\begin{align}
\mathcal{R}=  r_x(\alpha_3) r_y(\alpha_2) r_z(\alpha_1) ,
\end{align}
with
\begin{align}
r_z(\alpha_1)&= \left(
\begin{array}{ccccc}
\cos\alpha_1& \sin\alpha_1&0\\ 
-\sin\alpha_1&\cos\alpha_1& 0\\
0&0& 1
     \end{array}
     \right),
     \quad
r_y(\alpha_2) = \left(
\begin{array}{ccccc} 
\cos\alpha_2& 0& \sin\alpha_2\\
0&1 & 0 \\
-\sin\alpha_2 & 0&\cos\alpha_2 \\
  \end{array}
     \right),
   \nonumber \\
r_x(\alpha_3) &= \left(
\begin{array}{ccccc}  
1 & 0 & 0 \\
0&\cos\alpha_3& \sin\alpha_3 \\
0&-\sin\alpha_3& \cos\alpha_3 
         \end{array}
     \right).
\end{align}
where the mixing angles run over interval 0 to $\pi$.

In the  weak basis $( \phi_1\,, \phi_2\,,G\,,a)$,
the neutral scalar mass matrix ${\cal M}^2_0$ may be cast into
the following form~\cite{Pilaftsis:1999qt}:
\begin{equation}
  \label{NHiggs}
{\cal M}^2_0 \ =\ 
\left(\begin{array}{cc}  {\cal M}^2_S  & {\cal M}^2_{PS} \\
                {\cal M}^2_{SP} &  \widehat{\cal M}^2_P\end{array} \right)\, ,
\end{equation}
where $\widehat{\mathcal{M}}^2_P$  and $\mathcal{M}^2_S$ stand for the CP-conserving transitions $(G\,,a) \to (G\,,a)$ and $(\phi_1,\phi_2) \to (\phi_1,\phi_2)$,  respectively. The off-diagonal matrices ${\cal M}^2_{PS} = ({\cal
M}^{2}_{SP})^T$ contains the CP-violating mixings $a
\leftrightarrow (\phi_1,\phi_2)$. The elements of the above matrix
are given by
\begin{eqnarray}
\label{eq:M2S}
{\cal M}^2_S &=& M^2_a\, \left( \begin{array}{cc} 
 s^2_\beta & -s_\beta c_\beta \\ 
 -s_\beta c_\beta & c^2_\beta \end{array} \right)\, -\, 
\left( \begin{array}{cc} 
 \frac{\displaystyle T_{\phi_1}}{\displaystyle v\,c_\beta} & 0  \\ 
 0 & \frac{\displaystyle T_{\phi_2}}{\displaystyle v\,s_\beta} 
\end{array} \right)\ -v^2~ \left( \begin{array}{cc} 
X & Y  \\ 
Y & Z
\end{array} \right)\,,
\end{eqnarray}
with
\begin{eqnarray}
X&=& 2\lambda_1 c^2_\beta +  {\rm Re}(\lambda_5e^{2i\xi}) s^2_\beta
+ 2 {\rm Re}(\lambda_6 e^{i\xi}) s_\beta c_\beta\,,\nonumber \\
Y&=& 
(\lambda_3 + \lambda_4) s_\beta c_\beta + 
{\rm Re}(\lambda_6 e^{i\xi}) c^2_\beta + 
{\rm Re}(\lambda_7 e^{i\xi}) s^2_\beta\,, \nonumber \\
Z&=& 2\lambda_2 s^2_\beta +  {\rm Re}(\lambda_5 e^{2i\xi}) c^2_\beta
+ 2 {\rm Re}(\lambda_7 e^{i\xi}) s_\beta c_\beta\,.
\nonumber
\end{eqnarray}
In addition, we have
\begin{eqnarray}
  \label{eq:M2Phat}
\widehat{\cal M}^2_P & =&  \left( \begin{array}{cc} 
-\, \frac{\displaystyle c_\beta T_{\phi_1} + s_\beta
  T_{\phi_2}}{\displaystyle  v} 
& \frac{\displaystyle s_\beta T_{\phi_1} - c_\beta 
T_{\phi_2}}{\displaystyle v}\\
\frac{\displaystyle s_\beta T_{\phi_1} - c_\beta T_{\phi_2}}{\displaystyle v}&
\quad M^2_a\, -\, 
\frac{\displaystyle s_\beta\tan\beta\, T_{\phi_1} +
c_\beta\cot\beta\, T_{\phi_2}}{\displaystyle v} 
\end{array} \right)\,, \end{eqnarray}
\begin{eqnarray}
 \label{eq:M2SP}
{\cal M}^2_{PS} &=& v^2~ 
\left( \begin{array}{cc} 0 & \frac12{\rm Im}(\lambda_5 e^{2i\xi}) s_\beta\,
    +\, {\rm Im}(\lambda_6 e^{i\xi}) c_\beta \\
    0 & \frac12{\rm Im}(\lambda_5 e^{2i\xi}) c_\beta\,
    +\, {\rm Im}(\lambda_7 e^{i\xi}) s_\beta \end{array}\right)\, 
-\, \frac{T_a}{v} \left(\begin{array}{cc} 
s_\beta & c_\beta \\ - c_\beta & s_\beta \end{array} \right)\, ,
\end{eqnarray}
where
\begin{equation}
\label{SMa}
M^2_a\ =\ \frac{1}{2s_\beta c_\beta}\, 
\Big\{{\rm Re}( m^2_{12} e^{i\xi}) - v^2 \Big[ 
2{\rm Re}(\lambda_5 e^{2i\xi}) s_\beta c_\beta +
 \, {\rm Re}(\lambda_6 e^{i\xi}) c^2_\beta
+  \, {\rm Re}(\lambda_7 e^{i\xi}) s^2_\beta \Big]\, \Big\}\,.
\end{equation}
In the CP-conserving limit, $M_a$  becomes the mass of the CP-odd Higgs scalar. If we project out the zero mass-matrix entries pertinent to the neutral $G$-boson, the reduced neutral-scalar mass matrix in the basis $(\phi_1\,,\phi_2\,,a)$ becomes
\begin{align}
M^2=\left(
\begin{array}{ccccc}
M^2_{S_{11}} & {\cal M}^2_{S_{12}} & {\cal M}^2_{{PS}_{12}} 
\\
{\cal M}^2_{S_{21}} & {\cal M}^2_{S_{22}} &  {\cal M}^2_{{PS}_{22}}   
\\
 {\cal M}^2_{{PS}_{12}} &  {\cal M}^2_{{PS}_{22}} & \widehat{\cal M}^2_{P_{22}}  
\end{array}
\right)=\left(
\begin{array}{ccccc}
M_{11}^2 & M_{12}^2 & M_{13}^2 
\\
M_{12}^2 & M_{22}^2 & M_{23}^2   
\\
M_{13}^2 & M_{23}^2 & M_{33}^2  
\end{array}
\right).
\label{mass_mat}
\end{align}
Then, the three physical masses squared are obtained by diagonalising $M^2$,
\begin{align}
\widetilde{\mathcal{M}}^{2}\: =\: 
\mathcal{R}\, M^2\, \mathcal{R}^{\sf T}\,.
\label{mass}
\end{align}

To discuss SM Higgs alignment, it proves convenient to transform the CP-even scalar fields from a weak basis with a generic choice of vacua to the so-called Higgs basis, where only one Higgs doublet acquires the SM VEV. This can be achieved by virtue of a common orthogonal transformation that involves the mixing angle $\beta$.
\begin{equation}
\begin{pmatrix}
\mathcal{H}^0_{1}\\ \mathcal{H}^0_2 \\ \mathcal{H}^0_3
\end{pmatrix}\:
= \:
\mathcal{R}_z(\beta)
\begin{pmatrix}
\phi_1 \\
\phi_2 \\
a
\end{pmatrix},
\end{equation}
where the rotation matrix $\mathcal{R}_z(\beta)$ is
\begin{equation}
\mathcal{R}_z(\beta) = 
\begin{pmatrix}
\cos \beta   & \sin \beta & 0 
\\ 
-\sin \beta & \cos \beta& 0 
\\
 0 & 0& 1 
\end{pmatrix}.
\end{equation}

Therefore,  the composition of the above scalar mass eigenstate in terms of their respective weak  fields in the Higgs basis  can be obtained by the following orthogonal transformations
\begin{equation}
\begin{pmatrix}
H_1 \\ H_2 \\ H_3 
\end{pmatrix}
=
\mathcal{R} \mathcal{R}_z(\beta)^\top \,
\begin{pmatrix}
\mathcal{H}^0_{1}\\ \mathcal{H}^0_2 \\ \mathcal{H}^0_3
\end{pmatrix}.
\end{equation}
In the Higgs basis $(\mathcal{H}^0_{1}\,,\mathcal{H}^0_2\,, \mathcal{H}^0_3)$, the $3\times 3$ mass matrix of the neutral scalars is given by
\begin{align}
\footnotesize
{\widehat{M}}^2=\left(
\begin{array}{ccccc}
{\hat{M}}_{11}^2 &{\hat{M}}_{12}^2 & {\hat{M}}_{13}^2 
\\
{\hat{M}}_{12}^2 & {\hat{M}}_{22}^2 & {\hat{M}}_{23}^2   
\\
{\hat{M}}_{13}^2 & {\hat{M}}_{23}^2 & M_{33}^2  
\end{array}
\right).
\label{mass_mat-HB}
\end{align}
In this context, the elements of the mass matrix in the Higgs basis can be given as:
{\allowdisplaybreaks
\begin{subequations} 
    \label{mfc2}
\begin{align}
\footnotesize
\hat{M}_{11}^2&= {1\over 2} (M_{11}^2 + M_{22}^2 + (M_{11}^2 - M_{22}^2) \cos2\beta + 2 M_{12}^2 \sin2\beta),
 \\
\hat{M}_{22}^2&=  {1\over 2} (M_{11}^2 + M_{22}^2 + (-M_{11}^2 + M_{22}^2) \cos2\beta - 2 M_{12}^2 \sin2\beta),
\\
\hat{M}_{33}^2&=  M_{33}^2,
\\
\hat{M}_{12}^2&=  M_{12}^2 \cos2\beta - {1\over 2} (M_{11}^2 - M_{22}^2) \sin2\beta,
\\
\hat{M}_{13}^2&=  M_{13}^2 \cos\beta + M_{23}^2 \sin\beta,
  \\
\hat{M}_{23}^2&=  M_{23}^2 \cos\beta - M_{13}^2 \sin\beta.
  \end{align}
\end{subequations}}

 In the SM-like $H_1$-boson scenario of SM alignment, we may now perform a diagonalisation of the CP-even mass matrix ${\hat{M}}^2$ in two steps.  In the first step, we use two angles, e.g.~$\alpha_{1,2}$, to project out the mass eigenvalue of the SM-like Higgs state
\begin{align}
    \label{mass-wtomH}
\widetilde{\mathcal{M}}^{2}=
\left(
\begin{array}{ccc}
{M}_{H}^2 & 0 & 0
\\
0 & {B}_1&{C}
\\
0 & {C} & {B}_2
\\
\end{array}
\right).
\end{align}
 Here, the mixing angles $\alpha_{1,2}$ are given in terms of the mass-matrix elements in~\eqref{mass_mat-HB} by
 \begin{align}
   \label{alpha1}
\tan \ & 2(\beta-\alpha_{1})=\ \frac{2 {\hat{M}}_{12}^2}{{\hat{M}}_{11}^2-{\hat{M}}_{22}^2},\; 
\\
 \label{alpha2}
\tan \ & 2\alpha_{2}=\ \frac{4(\hat{M}_{11}^2-\hat{M}_{22}^2)\big[ {\hat{M}}_{13}^2 \cos(\beta-\alpha_1)-{\hat{M}}_{23}^2 \sin(\beta-\alpha_1)\big]}{(\hat{M}_{11}^2-\hat{M}_{22}^2)(\hat{M}_{11}^2+{\hat{M}}_{22}^2-2 {\hat{M}}_{33}^2)+ \big[({\hat{M}}_{11}^2-{\hat{M}}_{22}^2)^2+4\hat{M}_{12}^4\big]\cos 2(\beta-\alpha_1)}\;.
\end{align}

The second step of diagonalisation consists of bringing the matrix~$\widetilde{\mathcal{M}}^{2}$ in a fully diagonal form.
Thus, the mixing angle $\alpha$ can be evaluated in terms of the above mass parameters
as 
\begin{align}
\tan{2\alpha_3}\ = \ {2 {C} \over {B}_1- {B}_2}\,.
 \label{alpha3}
\end{align}
Additionally, the squared masses of the two heavy $H_2$ and $H_3$ bosons are: 
\begin{align}
    \label{h1,h2}
  M^2_{H_2,H_3} &=  {1 \over 2}\bigg[\,{B}_1 + {B}_2 \mp \sqrt{({B}_1- {B}_2 )^2+4 {{C}}^2}\,\bigg].
\end{align}
Note that, we consider the mass hierarchy $M_H \le M_{H_2} \le M_{H_3}$, whereas ${M}_{H_1}^2\simeq 125$~GeV is the SM-like Higgs boson mass.

On the other hand, the normalised Higgs-vector boson couplings may be written as~\cite{Pilaftsis:1999qt},
\begin{align}
g_{H_jVV} &= c_\beta \mathcal{R}_{j1}  + s_\beta\mathcal{R}_{j2}.
\label{ei}
\end{align}
with $j=1,2,3$. From~\eqref{ei}, the following normalised couplings are obtained:
\begin{align}
   \label{eq:ei2}
g_{H_1VV} &= \cos(\beta-\alpha_1) \cos\alpha_2, \nonumber\\
g_{H_2VV} &= \sin(\beta-\alpha_1) \cos\alpha_3-\cos(\beta-\alpha_1) \sin\alpha_2\sin\alpha_3,\\
g_{H_3VV} &= \sin(\beta-\alpha_1) \sin\alpha_3-\cos(\beta-\alpha_1) \sin\alpha_2\cos\alpha_3\,.\nonumber
\end{align}
These normalised couplings satisfy the following sum rule:
\begin{align}
g_{H_1VV}^2+g_{H_2VV}^2+g_{H_3VV}^2=1\,.
\end{align}
Therefore, combining the expressions in~\eqref{alpha1} and \eqref{alpha3}, one may obtain the Higgs-gauge boson couplings given in~\eqref{eq:ei2}.
Thus, one may obtain the degree of misalignment in the presence of CP-violating parameters.

With the aid of~\eqref{eq:ei2},  \eqref{alpha1},\eqref{alpha2},  \eqref{alpha3}, and after introducing the following two small parameters:
\begin{align}
X_1=\frac{ {\hat{M}}_{12}^2}{{\hat{M}}_{11}^2-{\hat{M}}_{22}^2}\;, \qquad 
X_2=\frac{ {\hat{M}}_{13}^2}{{\hat{M}}_{11}^2-{\hat{M}}_{33}^2}\;,
\label{X1X2}
\end{align}
we may expand the normalized gauge couplings up to the second order in $X_{1,2}$ as follows:
\begin{align}
g_{H_1VV} &= 1- {1\over 2}(X_1^2 +X_2^2)+\mathcal{O}(X^4), \\
g_{H_2VV} &= X_1+{2X_2 {\hat{M}}_{23}^2 \over (\hat{M}_{11}^2-\hat{M}_{22}^2)(1+4X_1^2)-(1-X_2^2)(\hat{M}_{11}^2+\hat{M}_{22}^2-2\hat{M}_{33}^2) }+\mathcal{O}(X^4),\\[3mm]
g_{H_3VV} &= \big[1-{(g_{H_1VV}^2+g_{H_2VV}^2)}\big]^{1/2}.
\end{align}
Observe that the alignment limit of a SM-like $H_1$ scenario, with $g_{H_1VV}=1$ and $g_{H_2VV}=g_{H_3VV}=0$, can be achieved when $\alpha_1=\beta$ and $\alpha_{2,3}=0$. In the absence of CP violation, we have $\alpha_{2,3}=0$, and so the requirement that $\alpha_1=\beta$ suffices to ensure SM alignment. 

In the present study, we adopt the Type-II 2HDM, in which the two Higgs doublets couple distinctively to fermions: the first scalar doublet couples to down-type quarks and leptons, whereas the second one couples to up-type quarks. 
The Yukawa Lagrangian of the Type-II 2HDM reads
\bea
-\mathcal{L}_Y \;=\; 
y_u^{mn} \overline{Q}_L^m \widetilde{\Phi}_2 u_R^n
\;+\;
y_d^{mn} \overline{Q}_L^{m} \Phi_1 d^n_R 
\;+\;
y_e^{mn} {\overline{L}_L^{m}} \Phi_1 e^n_R 
\;+\;
\mathrm{H.c.\,},
\eea
where $Q^m_L = \big( u^m_L\,, d^m_L\big)^{\sf T}$, $L^m_L = \big( \nu^m_L\,, e^m_L\big)^{\sf T}$ (with $m=1,2,3$), $\widetilde{\Phi}_2 = i\sigma^2 \Phi^*_2$ represents the hypercharge-conjugate of $\Phi_2$, and $\sigma^2$ is the second Pauli matrix. Hence, the neutral Yukawa couplings in this Type-II 2HDM are:
\bea
g_{\bar{d} d H_j} &=& \frac{m_{d }}{v c_\beta}\,\Big(\mathcal{R}_{j1}-is_\beta \mathcal{R}_{j3}\gamma_5\Big),\\
g_{\bar{u} u H_j}  &=& \frac{m_{u }}{v s_\beta}\, \Big(\mathcal{R}_{j2}-ic_\beta \mathcal{R}_{j3}\gamma_5\Big),\\
g_{\bar{l} l H_j}  &=& \frac{m_{l }}{v c_\beta}\,\Big(\mathcal{R}_{j1}-is_\beta \mathcal{R}_{j3}\gamma_5\Big),
\eea
where $\gamma_5$ is the standard matrix that anti-commutes with $\gamma^\mu$. To second order in the small parameters~$X_{1,2}$ given in~\eqref{X1X2}, the $H_1$ couplings to fermions may be 
approximated as follows:
\begin{align}
g_{\bar{d} d H_1} &= \frac{m_{d}}{v c_\beta} \left[g_{H_1VV}-is_\beta \left(X_2 - \frac{X_2^3}{6}\right) \gamma_5\right], \\
g_{\bar{u} u H_1} &= \frac{m_{u}}{v s_\beta} \left[-X_1 + \frac{X_1^3}{6}-ic_\beta \left(X_2 - \frac{X_2^3}{6}\right)\gamma_5\right], \\
g_{\bar{l} l H_1} &= \frac{m_{l}}{v c_\beta}\left[g_{H_1VV}-is_\beta \left(X_2 - \frac{X_2^3}{6}\right) \gamma_5\right].
\end{align}

\newpage
\setcounter{equation}{0}
\section{Higgs Alignment Condition in a General Basis}\label{app:HAL}

In Appendix~\ref{app:2.2}, we have seen that SM Higgs alignment takes its simplest form in the so-called Higgs basis, where
$\langle \Phi_1\rangle = (0,0)^{\sf T}$ and $\langle \Phi_2 \rangle = (0,v/\sqrt{2})^{\sf T}$, and $v$ is the VEV of the SM. 
In the Higgs basis, not only the CP-even scalar mass matrix ${\cal M}^2_S$ in~\eqref{eq:M2S} becomes diagonal, but also the CP-odd mass matrix ${\cal M}^2_P$ in~\eqref{eq:M2Phat} and the charged scalar mass matrix ${\cal M}^2_\pm$ given in Eq.~(2.19) of~\cite{Pilaftsis:1999qt}. Instead, the scalar-pseudoscalar mass matrix~${\cal M}^2_{SP} = {(\cal M}^2_{PS})^{\sf T}$ given in~\eqref{eq:M2SP} vanishes identically. Nevertheless, it would
be preferable if a criterion can be established so that SM Higgs alignment can be detected in a general scalar-field basis, beyond the Higgs basis. 

In order to derive a basis-independent SM-Higgs alignment condition, the following two Sp(4)-covariant quantities are introduced:
\begin{eqnarray}
   \label{eq:CovTad2}
  {\cal T}^{2,\,J}_{\, I} &=& C^{JK}\,\big< \Phi_I \big> \,\big< \Phi_K\big>\;,\\
  \label{eq:CovM2IJ}
  {\cal M}^{2,\,J}_{\, I} & = & C_{IK}\,\Big< \frac{\partial^2{\cal V}}{\partial \Phi_K \partial \Phi_J}  \Big>\;,
\end{eqnarray}
where $\{ \Phi_I\} = \bm{\Phi}$ (with $I = 1,2,3,4$) given in~\eqref{eq:8DPhi} is a 4D complex vector in the fundamental representation of Sp(4), and $\{C^{IJ}\} \equiv C = \{C_{IJ}\}$ is the charge conjugation matrix defined after~\eqref{eq:su2sp4}. We should bear in mind that $C_{IJ}$, together with its inverse $C^{IJ}$, plays the role of a metric, i.e.~one can use these to lower and raise field indices in the Sp(4)~space. After minimizing the 2HDM potential through the vanishing of the tadpole conditions in a general basis, the mixed rank-2 tensors, $\{ {\cal T}^{2,\,J}_{\,I}\} \equiv \bm{{\cal T}^2}$ and $\{ {\cal M}^{2\,,J}_{\, I}\} \equiv \bm{{\cal M}^2}$, can be consistently defined as stated in~\eqref{eq:CovTad2} and~\eqref{eq:CovM2IJ}. 

Under an Sp(4) transformation, $\bm{\Phi} \to \bm{\Phi}' = U\bm{\Phi}$ with $U\in \text{Sp}(4)$, the rank-2 tensors, $\bm{{\cal T}^2}$ and $\bm{{\cal M}^2}$, transform as follows:
\begin{equation}
  \label{eq:rank2}
 \bm{{\cal T'}^2}\: =\: U\, \bm{{\cal T}^2}\, U^\dagger\,,\qquad 
 \bm{{\cal M'}^2}\: =\: U\, \bm{{\cal M}^2}\, U^\dagger\,.
\end{equation}
Then, the general condition that dictates SM-Higgs alignment in any weak scalar basis is given by the vanishing of the commutator,
\begin{equation}
  \label{eq:HALcond}
[\bm{{\cal T}^2}\,, \bm{{\cal M}^2}]\: =\:  {\bf 0}\;.
\end{equation}
Notice that this condition remains invariant under general Sp(4) transformations, including $G_R = \text{SU}(2)_{\rm HF}$ rotations of the $\text{U}(1)_Y$-invariant reparameterisation group.

\vfill\eject

\setcounter{equation}{0}
\section{Electron Electric Dipole Moment at Two Loops}\label{app:eEDM}

The individual two-loop contributions to the electron EDM may be summarized as follows${}$~\cite{Abe:2013qla,Bowser-Chao:1997kjp,Ellis:2008zy,Oredsson:2019mni}:
{\small
\begin{align}
\raisebox{-0.4in}{\includegraphics[width=0.18\textwidth]{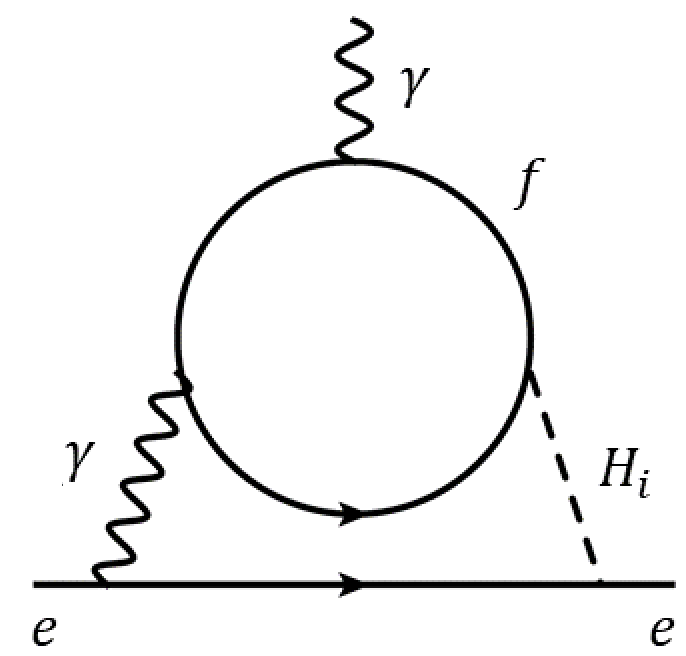}}: \frac{(d_e)^{\gamma H_i}_f}{e} &= \frac{\delta N_C e^2}{32 \pi^4} \sum_{f=\tau,t,b} \frac{Q_f^2}{m_f} \sum_{i=1}^3 \left[f\left(\frac{m_f^2}{M_{H_i}^2}\right)g_{H_i\bar{f}f} \tilde{g}_{H_i\bar{e}e}
\right.\nonumber \\ &\left. + g\left(\frac{m_f^2}{M_{H_i}^2}\right) g_{H_i\bar{e}e}~\tilde{g}_{H_i\bar{f}f}\right],
\label{A.1}
\\
\raisebox{-0.4in}{\includegraphics[width=0.18\textwidth]{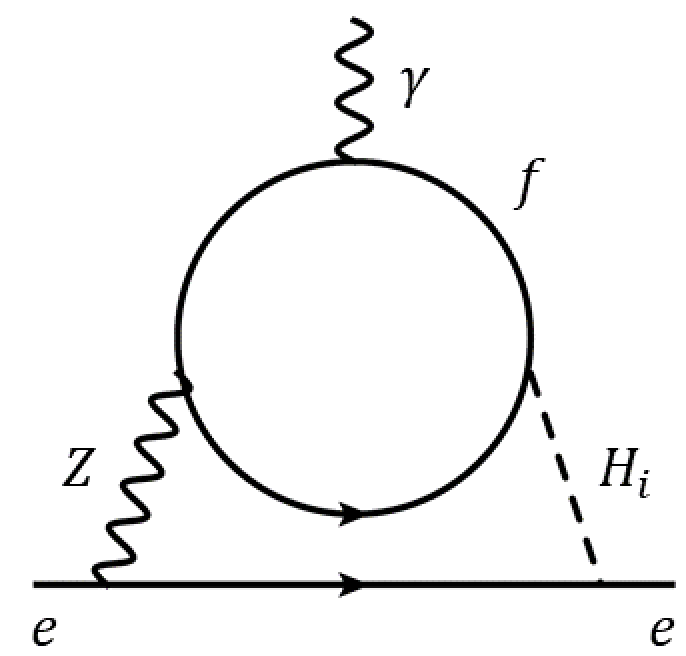}}: \frac{(d_e)^{ZH_i}_f}{e} &= \frac{\delta N_C g_{Zee}}{32 \pi^4} \sum_{f=\tau,t,b} \frac{Q_f g_{Zff}}{m_f} \sum_{i=1}^3\left[F\left(\frac{m_f^2}{M_{H_i}^2},\frac{m_f^2}{M_{Z}^2}\right)g_{H_i\bar{f}f} \tilde{g}_{H_i\bar{e}e}
\right.\nonumber \\ &\left. +G\left(\frac{m_f^2}{M_{H_i}^2},\frac{m_f^2}{M_{Z}^2}\right)g_{H_i\bar{e}e}~ \tilde{g}_{H_i\bar{f}f}\right],
\label{A.2}
\end{align}
}
\noindent
where $g_{Zff} = 2 e (T^3_f-2Q_f \sin^2\theta_W)/\sin 2\theta_W$ and $\delta = 1.96 \times 10^{-13}$ [cm$\cdot$GeV].
{\small
\begin{align}
\raisebox{-0.4in}{\includegraphics[width=0.18\textwidth]{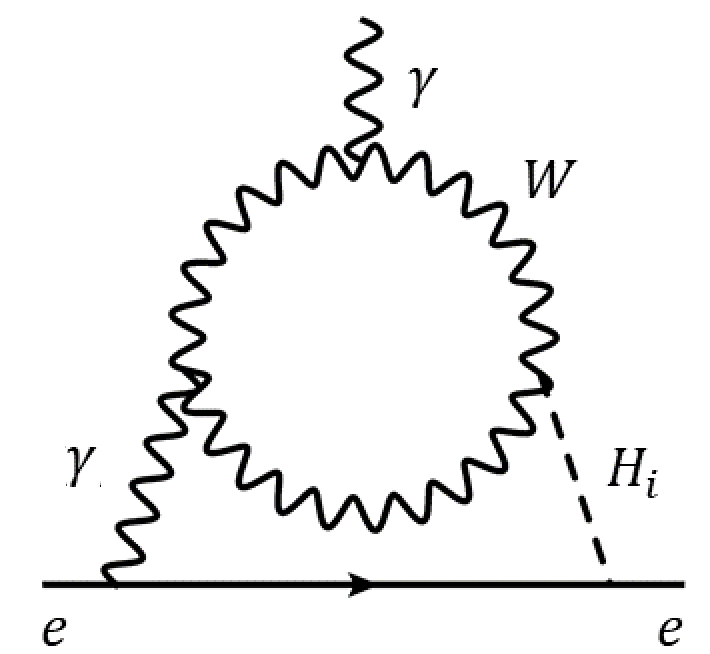}}: \frac{{(d_e})^{\gamma H_i}_W}{e} &= -\frac{\delta e^2}{128 \pi^4 v} \sum_{i=1}^3 \left[ \left(6+ \frac{M_{H_i}^2}{M_W^2}\right) f\left(\frac{M_W^2}{M_{H_i}^2}\right) 
\right.\nonumber \\ &\left.
+ \left(10- \frac{M_{H_i}^2}{M_W^2}\right) g\left(\frac{M_W^2}{M_{H_i}^2}\right)\right]
 g_{H_iVV}\, \tilde{g}_{H_i\bar{e}e},
\label{A.3}
\\
\raisebox{-0.4in}{\includegraphics[width=0.18\textwidth]{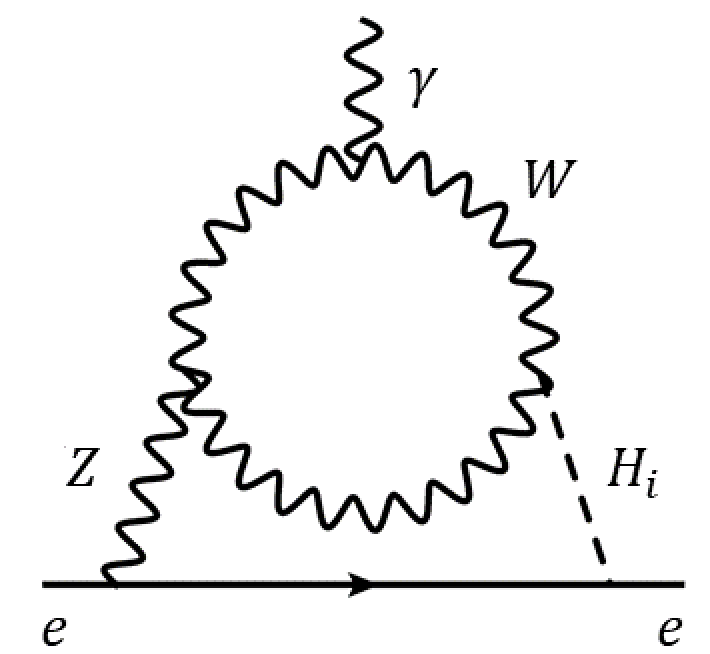}}: \frac{{(d_e})^{Z H_i}_W}{e} &= \frac{\delta e g_{Zee} g_{ZWW}}{128 \pi^4 \cos^4\theta_W v}\sum_{i=1}^3
\left[ \left(-1+6 \cos^2\theta_W- \frac{M_{H_i}^2}{2 M_W^2}(1-2 \cos^2\theta_W\right)
\right.\nonumber \\ &\left.
\times F\left(\frac{M_W^2}{M_{H_i}^2},\cos^2\theta_W\right)
- \left(3-10 \cos^2\theta_W+ \frac{M_{H_i}^2}{2 M_W^2}(1-2 \cos^2\theta_W)\right)
\right.\nonumber \\ &\left.
\times G\left(\frac{M_W^2}{M_{H_i}^2},\cos^2\theta_W\right)\right] g_{H_i VV} \tilde{g}_{H_i\bar{e}e},
\label{A.4}
\end{align}
}
with $g_{ZWW} = e \cot \theta_W$.
{\small
\begin{align}
\raisebox{-0.4in}{\includegraphics[width=0.18\textwidth]{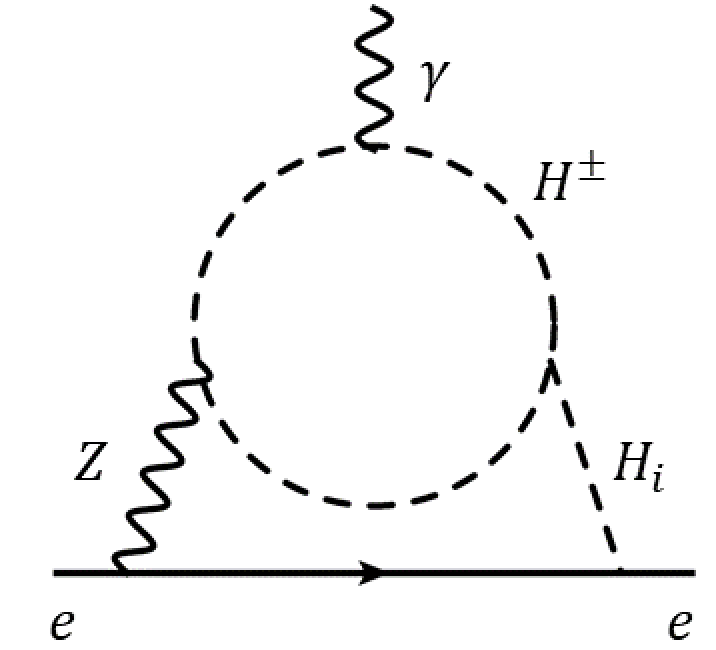}}: \frac{(d_e)^{Z H_i}_{H^\pm}}{e} &= -\frac{\delta g_{Zee} \,g_{ZH^\pm}}{128 \pi^4 M_{H^\pm}^2} \sum_{i=1}^3 \Bigg[F\left(\frac{M_{H^\pm}^2}{M_{H_i}^2},\frac{M_{H^\pm}^2}{M_{Z}^2}\right)- G\left(\frac{M_{H^\pm}^2}{M_{H_i}^2},\frac{M_{H^\pm}^2}{M_{Z}^2}\right)\Bigg]
\nonumber \\ & \times g_{H^\pm H_i}\tilde{g}_{H_i\bar{e}e},
\label{A.5}
\\
\raisebox{-0.4in}{\includegraphics[width=0.18\textwidth]{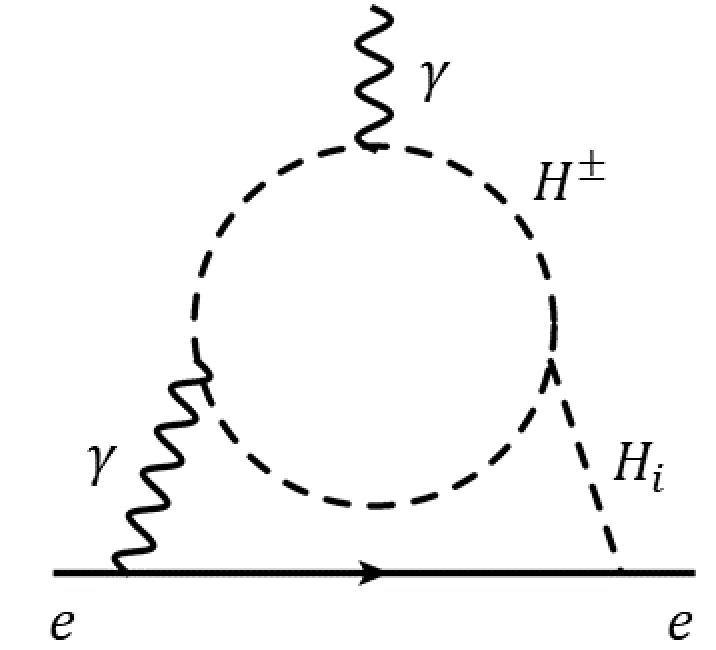}}: \frac{(d_e)^{\gamma H_i}_{H^\pm}}{e} &= -\frac{\delta e^2}{128 \pi^4 M_{H^\pm}^2} \sum_{i=1}^3 \Bigg[f\left(\frac{M_{H^\pm}^2}{M_{H_i}^2}\right)- g\left(\frac{M_{H^\pm}^2}{M_{H_i}^2}\right)\Bigg] g_{H^\pm H_i}\tilde{g}_{H_i\bar{e}e},
\label{A.6}
\end{align}
}
with $g_{ZH^\pm} = e \cot 2\theta_W$.
\\
{\small
\begin{align}
\raisebox{-0.4in}{\includegraphics[width=0.36\textwidth]{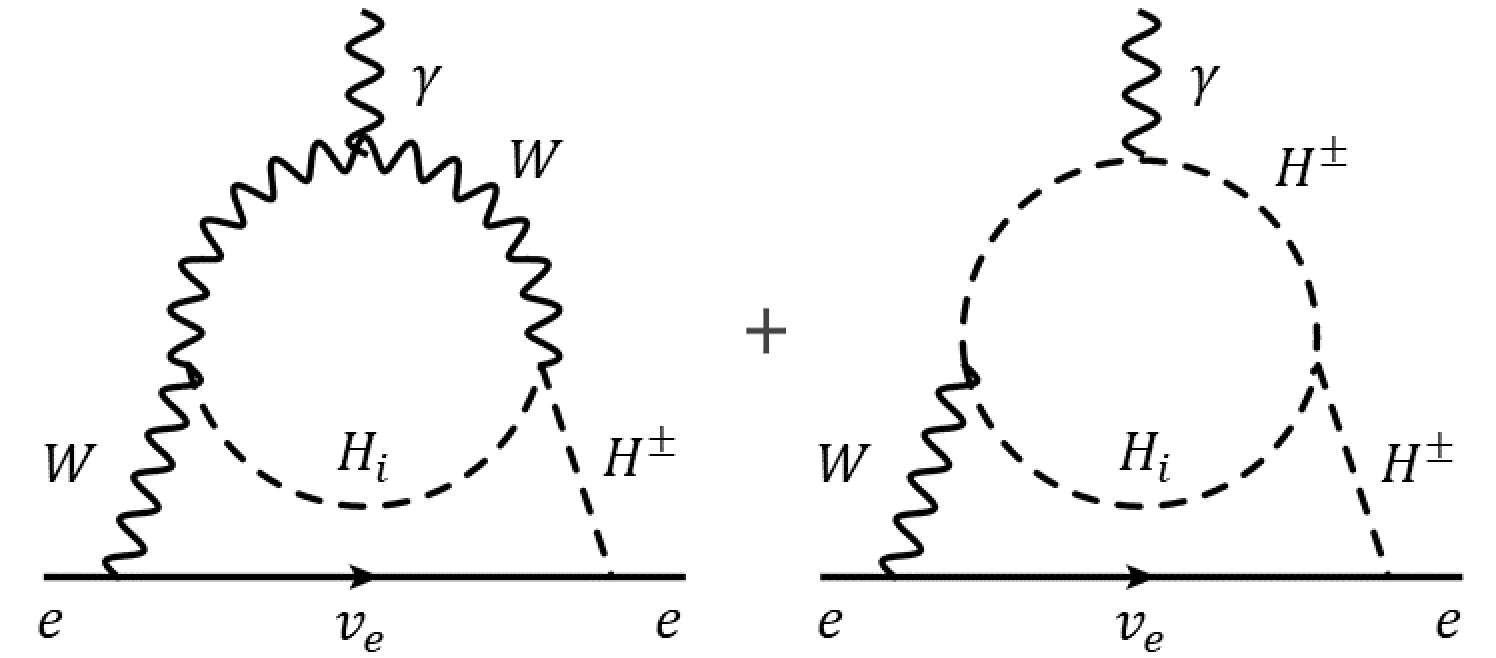}}: \frac{(d_e)^{H_i}_{W,H^\pm}}{e} &= \frac{\delta}{512 \pi^4 v^2} \sum_{i=1}^3 \Bigg[\frac{v g^2}{M_{H_i}^2} H\left(M_{H_i}^2,M_{H^\pm}^2\right) g_{H_iVV}
\nonumber \\ &
-2 G\left(M_{H_i}^2,M_{H^\pm}^2\right) g_{H^\pm H_i}\Bigg]\tilde{g}_{H^\pm\bar{\nu_e}e}.
\label{A.7}
\end{align}
}

\noindent
We note that the theoretical prediction for the electron EDM, $d_e$, is given by the sum of the individual two-loop contributions listed in~\eqref{A.1}--\eqref{A.7}.

In evaluating the above two-loop EDM effects, we should know 
the charged Higgs-boson~($H^\pm$) Yukawa couplings to quarks and leptons, in addition to the neutral Higgs couplings given in Appendix~\ref{app:2.2}. The $H^\pm$-boson couplings to fermions are given by 
\begin{align}
g_{\bar{l}_m\nu_mH^-} &= -\frac{m_{l_m}}{\sqrt{2}v}t_\beta(1-\gamma_5),\\
g_{\bar{\nu}_ml_mH^+} &= -\frac{m_{l_m}}{\sqrt{2}v}t_\beta(1+\gamma_5),\\
g_{\bar{u}_nd_mH^+} &= -
\frac{V_{nm}}{\sqrt{2}v c_\beta s_\beta}
\left[c_\beta^2m_{u_n}(1-\gamma_5)+s_\beta^2m_{d_m}(1+\gamma_5)\right],\\
g_{\bar{d}_mu_nH^-} &= -\frac{V_{nm}^*}{\sqrt{2}vc_\beta s_\beta}
\left[c_\beta^2m_{u_n}(1+\gamma_5)+s_\beta^2m_{d_m}(1-\gamma_5)\right],
\end{align}
where $V$ is the well-known CKM matrix describing the mixing between quarks~\cite{Kobayashi:1973fv}. Moreover,
trilinear couplings of neutral scalars to the charged scalar $H^{\pm}$ enter the $H_i \gamma \gamma$ coupling through the $H^{\pm}$ loop. These trilinear couplings are found to be
\begin{align}
g_{H_i H^+ H^-} &= \frac{1}{v^2} \Bigg[ \mathcal{R}_{i1} v_1 v_2^2 \Big(\lambda_1 - \lambda_4 - \mathrm{Re}(\lambda_5) + \frac{v_1^2}{v_2^2} \lambda_3 + \frac{v_2^2 - 2v_1^2}{v_1 v_2} \mathrm{Re}(\lambda_6) + \frac{v_1}{v_2} \mathrm{Im}(\lambda_6)\Big)
\nonumber \\
&\quad + \mathcal{R}_{i2} v_1^2 v_2 \Big(\lambda_1 - \lambda_4 - \mathrm{Re}(\lambda_5) + \frac{v_2^2}{v_1^2} \lambda_3 + \frac{v_1^2 - 2v_2^2}{v_1 v_2} \mathrm{Re}(\lambda_6) + \frac{v_2}{v_1} \mathrm{Im}(\lambda_6)\Big)
\nonumber \\
&\quad + \mathcal{R}_{i3} v \Big(v_1 v_2 \mathrm{Im}(\lambda_5) + v^2 \mathrm{Im}(\lambda_6)\Big) \Bigg],
\end{align}
with $i=1,2,3$ and ${\rm Im} (\lambda_5)=0$.

Finally, the loop functions used for the calculation of the electron EDM are~\cite{Barr:1990vd,Abe:2013qla}:
{ \small 
\begin{align}
f(z) &= \frac{z}{2}\int_0^1 dx \left[ \frac{1-2x(1-x)}{x(1-x)-z} \right] \log\left(\frac{x(1-x)}{z}\right),
\\
g(z) &= \frac{z}{2}\int_0^1 dx \left[ \frac{1}{x(1-x)-z} \right] \log\left(\frac{x(1-x)}{z}\right),
\\
F(x,y) &= \frac{1}{y-x}\left[y f(x) - x f(y) \right],
\\
G(x,y) &= \frac{1}{y-x} \left[y g(x) - x g(y) \right],
\\
H(m_1^2,m_2^2) &= \frac{M_{W}^2}{M_{H^\pm}^2-M_{W}^2}\left[ A\left(M_{W}^2, m_1^2\right) - A\left(m_2^2, m_1^2\right) \right],
\\
G(m_1^2,m_2^2) &= \frac{M_{W}^2}{M_{H^\pm}^2-M_{W}^2}\left[ B\left(M_{W}^2, m_1^2\right) - B\left(m_2^2, m_1^2\right) \right],
\end{align}
}
with
{\small
\begin{align}
A(m_1^2,m_2^2) &= \int_0^1 dz (1-z)^2 \left(z-4+z \frac{M_{H^\pm}^2-m_2^2}{M_{W}^2}\right)
\nonumber \\
&\quad \times \left[ \frac{m_1^2}{M_{W}^2(1-z)+m_2^2 z -m_1^2 z (1-z)} \right]\log\left(\frac{M_{W}^2 (1-z)+m_2^2 z}{m_1^2 z(1-z)}\right),
\\
B(m_1^2,m_2^2) &=  \int_0^1 dz \left[ \frac{m_1^2 z (1-z)^2}{M_{W}^2(1-z)+m_2^2 z -m_1^2 z (1-z)} \right]\log\left(\frac{M_{W}^2 (1-z)+m_2^2 z}{m_1^2 z(1-z)}\right).
\end{align}
}

\vfill\eject
\bibliography{biblio}
\end{document}